\documentclass[aps,prd,10pt,twocolumn,superscriptaddress,nofootinbib,floatfix]{revtex4-2}
\usepackage{amsmath,amssymb,graphicx,bm,placeins}
\usepackage[colorlinks=true,linkcolor=blue,citecolor=blue,urlcolor=blue]{hyperref}
\hypersetup{pdftitle={Universality between the O(N) nonlinear sigma model and the vector phi4 model in view of the perturbative renormalization group},pdfauthor={Abouzeid M. Shalaby}}
\newcommand{\ep}{\epsilon}
\newcommand{\ve}{\varepsilon}
\begin{document}

\title{Universality between the \texorpdfstring{$O(N)$}{O(N)} nonlinear sigma model and the vector \texorpdfstring{$\phi^4$}{phi4} model\texorpdfstring{\\}{ } in view of the perturbative renormalization group}

\author{Abouzeid M. Shalaby}
\email{amshalab@qu.edu.qa}
\affiliation{Department of Physics and Materials Sciences, College of Arts and Sciences, Qatar University, Al Tarfa, Doha 2713, Qatar}

\date{\today}

\begin{abstract}
The few orders known for the renormalization group functions of the $O(N)$ nonlinear sigma model in $d=2+\ep$ are commonly
regarded as too short to say anything quantitative about three dimensions, and hence about its universality with the $O(N)$ vector
$\phi^4$ model. We show that this view must be revised at large $N$. As for the $4-\ve$ series of the $\phi^4$ model, whose effective parameter $3/(N+8)$ decreases with $N$, the four-loop $2+\ep$ series is effectively an expansion in
$\ep/(N-2)$, which is the smaller of the two parameters for all $N>7$. Taken by itself, with no input from the $\phi^4$ model, it gives accurate exponents in $d=3$. Its two-loop sum, which optimal truncation selects for $N\le11$,
yields the closed form $\nu=(N-2)/(N-1)$, which agrees with Monte Carlo and nonperturbative renormalization group results within about
$1\%$ for all $N\ge6$ and within $0.1\%$ at $N=100$, and with the independently truncated seven-loop $4-\ve$ series of the $\phi^4$ model
within $2.6\%$ for $N\ge6$. This result, and its error estimate, use only the one-, two- and three-loop coefficients, which were computed directly from diagrams; it is therefore free of the large-$N$ information that enters the $2+\ep$ series at four loops, and the comparison of $\nu$ between the two models involves only the two-loop terms of each expansion. For the exponent $\eta$ the raw series is of no use: its terms do not decrease at $\ep=1$ and its four-loop sum is negative
for every $N$, a fact we trace to the unit radius of convergence of the exact $O(1/N)$ result. Hypergeometric-Meijer resummation of the
same four coefficients restores the information. Since the large-order behavior of the $2+\ep$ series is unknown and Gevrey-1 growth is not guaranteed, we use both the divergent-type ${}_2F_0$ and the convergent-type ${}_2F_1$; the latter is favored by the geometric growth of the exact $O(1/N)$ and $O(1/N^2)$ coefficients, and it reproduces the exact $O(1/N^2)$ coefficient of $\eta$ in $d=3$ within $6\%$. The ${}_2F_1$ gives $\eta$ within $5\%$ for $6\le N\le12$, the ${}_2F_0$ within $4\%$ at $N=100$, and for every $N\ge8$ the two bracket the reference value.
For small $N$, where $\ep/(N-2)$ is of order one, four loops are not sufficient. Within perturbative renormalization group theory alone, the
expansions about two and four dimensions thus agree in the large-$N$ regime, consistent with the universality of the two models there. For $\nu$ this agreement is an independent test; for $\eta$, which requires the four-loop coefficient, it is partly built in, because that coefficient was fixed with the help of large-$N$ results.
\end{abstract}

\maketitle

\section{Introduction}
Continuum field theories emerge from very different microscopic descriptions, and the renormalization group explains why:
models with the same symmetry and dimensionality flow to the same fixed point and share its critical exponents
\cite{WilsonKogut,Wilson83,ZJbook,PV}. For an $N$-component order parameter with $O(N)$ symmetry there are two standard
continuum formulations. The first is the linear $\phi^4$ model, whose upper critical dimension is four and whose critical
point is the Wilson-Fisher (WF) fixed point, accessible in an expansion in $\ve=4-d$ \cite{WF}. The second is the nonlinear sigma
model (NLSM), in which the field is constrained to the sphere $S^{N-1}$. It is asymptotically free in two dimensions for $N>2$
\cite{Polyakov,ZamZam} and has a nontrivial ultraviolet fixed point in $d=2+\ep$ \cite{BZJ,HB,BW}. The universality hypothesis
identifies this fixed point, continued to $d=3$, with the WF one. The $O(N)$ universality classes are of wide physical interest: $N=2$ describes
the superfluid transition \cite{Campostrini06}, $N=3$ isotropic magnets, and $N=4$ the finite-temperature chiral transition of two-flavor
QCD \cite{PW}.

Numerically the identification is well supported. Lattice Monte Carlo studies of $O(N)$ spin and $\phi^4$ models give consistent
exponents, e.g., for $N=2$ \cite{Has19}, $N=4$ \cite{Deng}, and $N=6$ \cite{HS} (see Ref.~\cite{HV} for a compilation for $N=2$, $3$, $4$), and a recent qubit regularization of the NLSM, with an $(N+1)$-dimensional local Hilbert space, gave
$\nu$ and $\eta$ in agreement with the WF values up to $N=8$ \cite{Singh}. These can be compared with the conformal bootstrap
\cite{Chester,Kos16,Kos15}, with the $1/N$ expansion \cite{MZJ,VPK}, with the nonperturbative renormalization group (NPRG) \cite{DePolsi},
and with high-precision Monte Carlo (MC) results for $N\ge4$ \cite{Has}.

Analytically the situation is less clear. A functional renormalization group study of the NLSM in $d=3$ finds the expected fixed point only in a reduced
truncation, a fourth-order derivative operator destabilizing it \cite{FWZ}; this echoes the old stability problem of high-gradient
operators in $2+\ep$ dimensions, for which $1/N^2$ calculations favor the conventional fixed point \cite{DM}. More recently it
was pointed out that the NLSM fixed point in $2+\ep$ has a protected operator of dimension $N-1$ with no counterpart at the WF
fixed point, so that the two may fail to be analytically connected at finite $N$ unless multiplet recombination lifts it \cite{DCR}.
Quantitative tests that use only the $2+\ep$ data are therefore of interest.

On the perturbative side the two expansions have very different reputations. The $4-\ve$ series is known to seven loops
\cite{Schnetz,S751}; it is sign alternating and Borel summable, and yields precise exponents after resummation \cite{KP,HM1,S1049}. The $2+\ep$ series is known to
four loops \cite{HB,BW}, with partial information at five loops \cite{KSV}. For $\nu^{-1}$ it has no sign alternation, which was recognized early as a sign of non-Borel
summability \cite{HB}, and its direct use at $\ep=1$ has been called useless \cite{Kleinert}. Kleinert showed that the two series can be joined by
treating the $2+\ep$ expansion as the strong-coupling limit of the $4-\ve$ one, obtaining $\nu=0.695(10)$, $0.735(10)$, and $0.766(10)$ for $N=3,4,5$
\cite{Kleinert}; this, however, assumes the connection that one would like to test.

Previous perturbative work on the relation between the two models falls into three groups. (a) Consistency checks through the $1/N$ expansion, which is valid for all $2<d<4$: the coefficients of both $\ep$ expansions agree with the large-$N$ exponents order by order in $1/N$ \cite{BW,KSV}, and this agreement is commonly taken as the perturbative evidence that the fixed point can be followed smoothly from $d=4$ down to $d=2+\ep$, a conclusion drawn for the ferromagnetic $O(N)$ case and extended to helimagnets in Ref.~\cite{ADDJ}. It is an identity between coefficients at each order in $1/N$ and makes no statement at fixed finite $N$. (b) Interpolations that use both series as input \cite{Kleinert}. (c) Analyses of the $2+\ep$ expansion by itself, which either found it numerically poor at small $N$ \cite{HB,Kleinert,DCR} or questioned its validity, because of the infinite set of high-gradient operators \cite{CC,DM}, of the nonanalyticity predicted near $d=2$, $N=2$ \cite{CH} (not found for $d>2$ in a functional renormalization group study \cite{CJ}), or of the operator content \cite{DCR}. In a study of the operator content of the two fixed points, De Cesare and Rychkov also compared Pad\'e-resummed four-loop $2+\ep$ and $4-\ve$ results for 3D scaling dimensions with bootstrap data: the agreement is poor for $N=3$ and becomes reasonable for $N\gtrsim10$ \cite{DCR}. We are not aware of a study in which the $2+\ep$ and $4-\ve$ expansions are evaluated separately in $d=3$ and compared systematically as functions of $N$, for $\nu$ and $\eta$, against each other and against modern nonperturbative results. This is what we do here, and it makes the large-$N$ observation of Ref.~\cite{DCR} quantitative.

The prevailing view is that the four known orders of the $2+\ep$ functions are not sufficient for a quantitative study of the three-dimensional critical point \cite{HB,Kleinert}, so that the universality of the two models cannot be examined from the $2+\ep$ side within perturbation theory. The central point of this paper is that this view must be revised at large $N$. Every correction in the $2+\ep$ series carries at least one power of $1/(N-2)$, so that the series is effectively an expansion in $\ep/(N-2)$. This parallels the $4-\ve$ series of the $\phi^4$ model, whose effective parameter $\sigma=3/(N+8)$ decreases with $N$, so that its resummed exponents become more accurate as $N$ grows \cite{S1049}. The $2+\ep$ parameter decreases faster: $1/(N-2)$ is smaller than $3/(N+8)$ for all $N>7$, and almost three times smaller at $N=100$. As $N$ grows the $2+\ep$ series therefore becomes accurate by itself, and where the raw series fails, as for $\eta$, hypergeometric-Meijer resummation of the same coefficients restores accurate values.

In Ref.~\cite{S751} it was shown that for $N\ge4$ the $4-\ve$ series needs no resummation: its optimal truncation, i.e., the
partial sum that stops just before the smallest term, already gives $\nu$ and $\eta$ at a level competitive with MC,
bootstrap and NPRG results. Here we apply the same rule to the four-loop $2+\ep$ series of the NLSM and compare the two optimally
truncated expansions with each other and with nonperturbative results. This is a test within perturbative renormalization group theory that uses nothing beyond the raw
coefficients of each expansion separately: the two series are never combined, and neither is used to constrain the other. The coefficients themselves, however, are not entirely free of large-$N$ input: the terms through three loops were computed directly from diagrams \cite{BZJ,HB}, but the four-loop $\beta$ function, which enters the $\ep^4$ coefficients of both $\nu^{-1}$ and $\eta$, was fixed with the help of the $1/N$ expansion of $\eta$ \cite{BW,Wegner89}. The result for $\nu$ is unaffected by this: the two-loop truncation and its error estimate use only the directly computed terms, and so does the two-loop truncation on the $\phi^4$ side, so that the comparison of $\nu$ between the two models is an independent test. For $\eta$ the four-loop coefficient is indispensable (Sec.~\ref{sec:disc}), and there the agreement at large $N$ is partly built in. What the comparison measures is how well each truncated series, at fixed finite $N$, reproduces the three-dimensional values. It yields a closed-form two-loop expression from each side, Eq.~(\ref{nuOT}) from $d=2$ and Eq.~(\ref{phi4OT}) from $d=4$, whose agreement or disagreement can be read off directly. We complement it with the lowest-order hypergeometric-Meijer resummation of the same
coefficients and with an analysis based on the exact large-$N$ results.

Our main results can be summarized as follows; the figures referred to illustrate them.
(i) The terms of the $2+\ep$ series of $\nu^{-1}$ decrease up to the three-loop one for $N\le11$ [Fig.~\ref{figterms}], so that the optimally truncated
series is the two-loop sum, which gives the closed form $\nu=(N-2)/(N-1)$; its partial sums are shown in Fig.~\ref{figpartial}. This value and its error estimate rest only on the directly computed one-, two- and three-loop coefficients, with no large-$N$ input.
(ii) The terms of the $\eta$ series never decrease at $\ep=1$, and the partial sums oscillate [Fig.~\ref{figpartial}(b)]; we trace this to the unit radius of convergence of the exact
$O(1/N)$ result [Fig.~\ref{figlargeN}(a)].
(iii) The accuracy of the $2+\ep$ series is controlled by the effective parameter $\ep/(N-2)$, which is smaller than the corresponding $4-\ve$ parameter for $N>7$ [Fig.~\ref{figlargeN}(b)].
(iv) Hypergeometric-Meijer resummation restores $\eta$; a convergent-type ${}_2F_1$ is favored by the available higher-order information, and together with the
divergent-type ${}_2F_0$ it brackets the reference value of $\eta$ for $N\ge8$. The exact $O(1/N^2)$ exponents, which have unit radius of convergence in $\ep$ for both $\eta$ and $\nu^{-1}$,
confirm the geometric growth and show that the ${}_2F_1$ extrapolates correctly at that order [Table~\ref{tabNexp}].
(v) The deviations from nonperturbative results decrease with $N$ for both exponents [Fig.~\ref{figdev}], and the expansions about two and four dimensions agree in the
large-$N$ regime.

The paper is organized as follows. Section~\ref{sec:ot} recalls the optimal truncation rule. Sections~\ref{sec:nu} and \ref{sec:eta}
apply it to $\nu$ and $\eta$. Section~\ref{sec:largeN} interprets the results through the exact $O(1/N)$ and $O(1/N^2)$ exponents and comments on the
large-order behavior. Section~\ref{sec:hm} presents the hypergeometric-Meijer resummation, Sec.~\ref{sec:comp} the comparison of the two models,
Sec.~\ref{sec:disc} a discussion of what the results imply for universality and of their limitations, and Sec.~\ref{sec:concl} our conclusions.
Appendix~\ref{app:mb} gives the Mellin-Barnes evaluation of the approximants.

\section{Optimal truncation}\label{sec:ot}
A series $\sum_i c_i\alpha^i$ is asymptotic to a function $f(\alpha)$ if, for every $K$, the remainder $f-\sum_{i\le K}c_i\alpha^i$ behaves as $c_M\alpha^M$, with $c_M$ the first
nonvanishing coefficient beyond $c_K$ \cite{BO}. For factorially growing coefficients the terms decrease to a minimum and then grow. The optimal
truncation rule is to stop just before the smallest term; the error is then of the order of that term \cite{BO,S751}.
For the $4-\ve$ series the terms behave as $i!\,(-\sigma\ve)^i$ with $\sigma=3/(N+8)$, so that the order of the minimum grows like
$1/(\sigma\ve)$ and the truncation becomes more accurate as $N$ increases \cite{S751}.

We denote by $T_k$ the $k$th term at $\ep=1$ and by $S_k=\sum_{j\le k}T_j$ the partial sums. Two conditions are needed for the rule to be meaningful.
(i) The terms must decrease initially. (ii) The smallest term must be a minimum of the sequence at fixed $N$ produced by the
growth of the coefficients, and not a term that is small because its coefficient passes through zero as a function of $N$. Both conditions
turn out to matter below.

The rule rests on the structure of asymptotic series. If the coefficients grow as $c_i\sim i!\,A^{-i}$ (Gevrey-1 growth), the term
$c_i\alpha^i$ is smallest at $i\simeq A/\alpha$, and the error of the truncated sum is of the order of $e^{-A/\alpha}$, which is the size of the
nonperturbative ambiguity of the function itself \cite{BO}. Optimal truncation thus extracts all the information the series contains, without
assuming anything about its resummation. For a convergent series the terms decrease indefinitely and the rule simply says to keep all known
terms. The rule therefore has a definite meaning only when the known terms show a minimum, which is what conditions (i) and (ii) test.
With four terms, as here, a minimum can be identified only if it falls at the second, third or fourth order, and the corresponding error
estimate, the first omitted term, is necessarily crude. We nevertheless find below that, where the conditions hold, it is a faithful measure of the
actual error.

\section{The exponent \texorpdfstring{$\nu$}{nu}}\label{sec:nu}
\begin{figure*}[tbp]\centering
\includegraphics[width=0.86\textwidth]{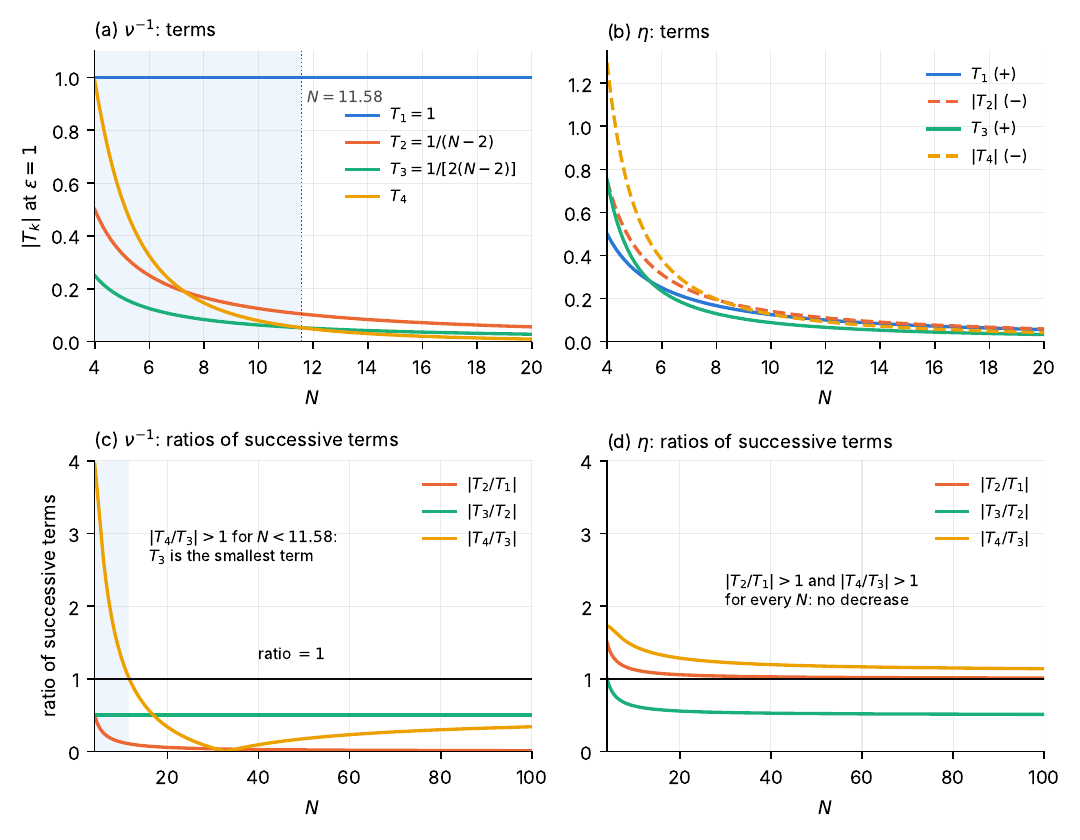}
\caption{Terms $T_k$ of the four-loop $2+\ep$ series at $\ep=1$ ($d=3$) as functions of $N$. Top: magnitudes $|T_k|$ for $4\le N\le20$; dashed lines mark negative terms. Bottom: ratios of successive terms $|T_{k+1}/T_k|$ for $4\le N\le100$; a ratio above $1$ means that the next term is larger.
(a), (c) $\nu^{-1}$, Eq.~(\ref{nu}): the terms first decrease ($|T_2/T_1|<1$, $|T_3/T_2|=1/2$). For $N<11.58$ (shaded) $|T_4/T_3|>1$, so that $T_3$ is the smallest term and the optimal truncation is the two-loop sum, Eq.~(\ref{nuOT}).
For $N>11.58$ the ratio $|T_4/T_3|$ falls below $1$, and it vanishes at $N=32.74$ where the four-loop coefficient changes sign.
(b), (d) $\eta$, Eq.~(\ref{eta}): the terms alternate in sign and have comparable magnitudes. $|T_2/T_1|=(N-1)/(N-2)$ and $|T_4/T_3|$ exceed $1$ for every $N$, so that the terms do not decrease and there is no minimum produced by the growth of the coefficients.
All ratios tend to constants as $N\to\infty$ while the terms themselves fall off like $1/N$, which reflects the effective parameter $\ep/(N-2)$ of Sec.~\ref{sec:largeN}.}
\label{figterms}
\end{figure*}

With $M=N-2$ and $d=2+\ep$, the four-loop series reads \cite{HB,BW,Kleinert}
\begin{align}
\nu^{-1}&=\ep+\frac{\ep^2}{M}+\frac{\ep^3}{2M}+c_3\,\ep^4+O(\ep^5),\label{nu}\\
c_3&=-\frac{30-14N+N^2+(54-18N)\zeta(3)}{4M^3}\nonumber\\
&=-\frac{1}{4M}+\frac{7.9093}{M^2}-\frac{6.9093}{M^3}.\nonumber
\end{align}
At $\ep=1$ the terms are $T_1=1$, $T_2=1/M$, $T_3=1/(2M)$, and $T_4=c_3$. They are all positive for $N<32.74$, where $c_3$ changes sign.

For $3\le N\le 11$ one has $T_4>T_3$ (the crossing is at $N=11.58$). The smallest term is then the three-loop one, and the optimal
truncation is the two-loop sum,
\begin{equation}
\nu^{-1}\simeq 1+\frac{1}{N-2},\qquad \nu_{\rm OT}=\frac{N-2}{N-1},
\label{nuOT}
\end{equation}
with the error estimate $T_3=1/[2(N-2)]$ on $\nu^{-1}$. The terms are listed in Table~\ref{tabterms} and shown in Fig.~\ref{figterms}(a), and the results are collected in Table~\ref{tabnu}, where the columns are grouped by model.
For $N=4$ to $10$ the reference value of $\nu^{-1}$ lies inside the band $S_2\pm T_3$ in every case, e.g., $1.25\pm0.125$ against the
MC value $1.2375(9)$ at $N=6$ \cite{Has}. The deviation of $\nu_{\rm OT}$ from the reference is $-10.9\%$, $-3.9\%$, $-1.0\%$, $+0.7\%$, and $+1.0\%$ for
$N=4,5,6,8,10$. At $N=3$ the first two terms are equal and the rule gives $\nu=0.5$ against $0.7073(30)$ \cite{Kleinert}; the natural parameter
$\ep/(N-2)$ equals one there and the truncation cannot be expected to work.

\begin{figure*}[tbp]\centering
\includegraphics[width=0.86\textwidth]{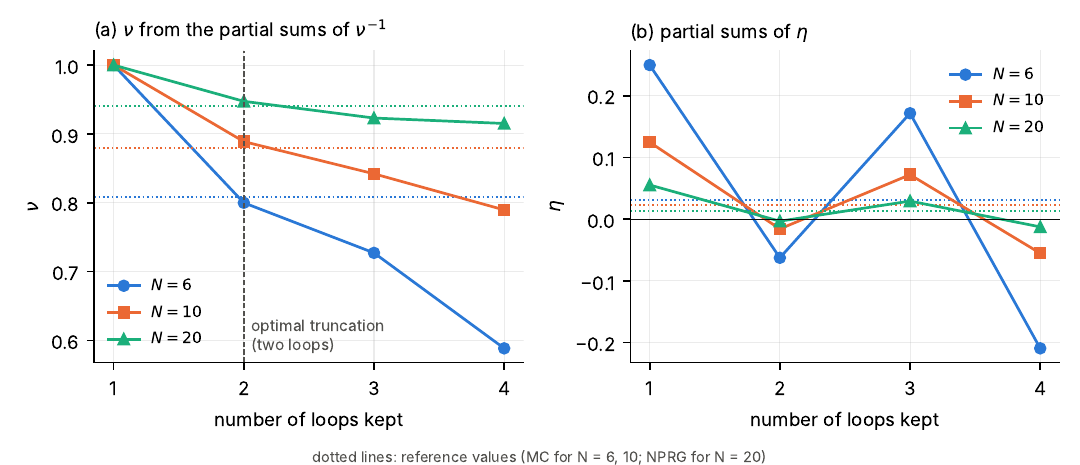}
\caption{Partial sums of the four-loop $2+\ep$ series at $\ep=1$ against the number of loops kept, for $N=6$, $10$, and $20$; dotted lines are the reference values of
Tables~\ref{tabnu} and \ref{tabeta}. (a) $\nu=1/S_k$ from the partial sums of $\nu^{-1}$: the two-loop sum, selected by optimal truncation for $N\le11$, is the closest to
the reference for all three values of $N$, and adding the three- and four-loop terms moves the result away. (b) Partial sums of $\eta$: they alternate in sign around the
reference value without settling; beyond two loops each added term moves the sum further from the reference, and the four-loop sum is negative for every $N$.}
\label{figpartial}
\end{figure*}

The partial sums are displayed in Fig.~\ref{figpartial}(a). For $N=6$, $10$, and $20$ the sequence $1/S_k$ is $1$, $0.800$, $0.727$, $0.589$; $1$, $0.889$, $0.842$, $0.790$;
and $1$, $0.947$, $0.923$, $0.915$, against the reference values $0.808$, $0.880$, and $0.941$. In each case the two-loop sum is the closest, and the
higher terms move the result away, which is the behavior expected below the smallest term of an asymptotic series.

The closed form~(\ref{nuOT}) has a transparent structure. Written as $\nu=1-1/(N-1)=1-1/N-1/N^2-\dots$, it has the form of the large-$N$ expansion
of the exact exponent, $\nu=1-1.0808/N+\dots$ in $d=3$, with a leading coefficient $1$ instead of $32/(3\pi^2)=1.0808$, i.e., $7.5\%$ too small. Since this coefficient multiplies $1/N$,
the resulting error in $\nu$ is about $0.08/N$: $0.0008$ at $N=100$, close to the actual deviation of $0.0011$. The two-loop truncation thus captures the
leading large-$N$ physics with a coefficient that is nearly right, which explains both its accuracy at large $N$ and the way that accuracy improves with $N$.

We stress that Eq.~(\ref{nuOT}) is free of large-$N$ input. Its two terms, and the three-loop term $T_3$ that serves as its error estimate, come from the direct one-, two- and three-loop calculations \cite{BZJ,HB}; the large-$N$ information used to fix the four-loop $\beta$ function \cite{BW,Wegner89} enters only $T_4$. The value of $\nu$, its error band $S_2\pm T_3$, which contains the reference value for every $N\ge4$ in Table~\ref{tabnu}, and its agreement with the two-loop $4-\ve$ truncation, Eq.~(\ref{phi4OT}), therefore owe nothing to the $1/N$ expansion. The four-loop term plays a role only in the choice of the truncation order: it is $T_4>T_3$ that marks the two-loop sum as optimal for $N\le11$; with three terms alone, which decrease monotonically, the rule would keep all three, and the three-loop sum is less accurate ($-10\%$ at $N=6$).

\begin{table*}[tbp]
\caption{Terms $T_k$ of the four-loop $2+\ep$ series of the nonlinear sigma model at $\ep=1$ ($d=3$): Eq.~(\ref{nu}) for $\nu^{-1}$, where $T_1=1$, and Eq.~(\ref{eta}) for $\eta$.
The smallest term in each series is in boldface; it fixes the optimal truncation.}
\label{tabterms}
\begin{ruledtabular}
\begin{tabular}{rccccccc}
 & \multicolumn{3}{c}{$\nu^{-1}$} & \multicolumn{4}{c}{$\eta$}\\
\cline{2-4}\cline{5-8}
$N$ & $T_2$ & $T_3$ & $T_4$ & $T_1$ & $T_2$ & $T_3$ & $T_4$\\
\hline
4 & 0.5000 & \textbf{0.2500} & 0.9887 & \textbf{0.5000} & $-0.7500$ & 0.7500 & $-1.2945$\\
5 & 0.3333 & \textbf{0.1667} & 0.5396 & \textbf{0.3333} & $-0.4444$ & 0.3704 & $-0.6251$\\
6 & 0.2500 & \textbf{0.1250} & 0.3239 & 0.2500 & $-0.3125$ & \textbf{0.2344} & $-0.3812$\\
8 & 0.1667 & \textbf{0.0833} & 0.1460 & 0.1667 & $-0.1944$ & \textbf{0.1296} & $-0.1973$\\
10 & 0.1250 & \textbf{0.0625} & 0.0788 & 0.1250 & $-0.1406$ & \textbf{0.0879} & $-0.1273$\\
12 & 0.1000 & 0.0500 & \textbf{0.0472} & 0.1000 & $-0.1100$ & \textbf{0.0660} & $-0.0922$\\
20 & 0.0556 & 0.0278 & \textbf{0.0093} & 0.0556 & $-0.0586$ & \textbf{0.0326} & $-0.0418$\\
32 & 0.0333 & 0.0167 & \textbf{0.0002} & 0.0333 & $-0.0344$ & \textbf{0.0184} & $-0.0224$\\
100 & 0.0102 & 0.0051 & $\mathbf{-0.0017}$ & 0.0102 & $-0.0103$ & \textbf{0.0053} & $-0.0060$\\
\end{tabular}
\end{ruledtabular}
\end{table*}

\begin{table*}[tbp]
\caption{Exponent $\nu$ in $d=3$, grouped by model. \emph{Nonlinear sigma model:} two-loop sum $S_2$, i.e.\ $\nu=(N-2)/(N-1)$; three-loop sum $S_3$, i.e.\ $\nu=[1+3/(2(N-2))]^{-1}$;
${}_2F_0$ resummation of Sec.~\ref{sec:hm} (median of the lateral sums for $N\le12$); these three use the four-loop $2+\ep$ series. Boldface marks the sum selected by the literal optimal truncation rule.
The fourth entry is the nonperturbative qubit-regularized model of Ref.~\cite{Singh}. \emph{$\phi^4$ model:} optimal truncation of the seven-loop $4-\ve$ series \cite{S751} and its hypergeometric resummation \cite{S1049}.
\emph{Reference:} MC \cite{Has} (for $N=6,8,12$ converted from the quoted $\nu^{-1}$), NPRG \cite{DePolsi}, and strong coupling (SC) \cite{KSF,KleinertSC}.}
\label{tabnu}
\begin{ruledtabular}
\begin{tabular}{rcccccccc}
 & \multicolumn{4}{c}{$O(N)$ nonlinear sigma model} & \multicolumn{2}{c}{$O(N)$ $\phi^4$ model} & \multicolumn{2}{c}{Reference}\\
\cline{2-5}\cline{6-7}\cline{8-9}
 & \multicolumn{3}{c}{$2+\ep$ expansion (four loops)} & Lattice & \multicolumn{2}{c}{$4-\ve$ expansion (seven loops)} & & \\
$N$ & Opt.\ trunc.\ $S_2$ & $S_3$ & ${}_2F_0$ resum. & Qubit MC & Opt.\ trunc. & Resummed & Value & Method\\
\hline
4 & \textbf{0.6667} & 0.5714 & 0.6778 & 0.7510(50) & 0.7500 & 0.7444(67) & 0.74817(20) & MC\\
5 & \textbf{0.7500} & 0.6667 & 0.7564 & & 0.7765 & 0.780(5) & 0.7802(6) & MC\\
6 & \textbf{0.8000} & 0.7273 & 0.8011 & 0.8040(40) & 0.7996 & & 0.8081(6) & MC\\
8 & \textbf{0.8571} & 0.8000 & 0.8483 & 0.8350(60) & 0.8373 & & 0.8509(7) & MC\\
10 & \textbf{0.8889} & 0.8421 & 0.8698 & & 0.8663 & 0.8792(9) & 0.8797(9) & MC\\
12 & 0.9091 & \textbf{0.8696} & 0.8767 & & 0.8889 & & 0.9003(14) & MC\\
20 & 0.9474 & \textbf{0.9231} & 0.9151 & & 0.9417 & 0.94784(52) & 0.9409(6) & NPRG\\
32 & 0.9677 & \textbf{0.9524} & 0.9523 & & 0.9729 & & 0.964 & SC\\
100 & 0.9899 & \textbf{0.9849} & 0.9858 & & & 0.9885(2) & 0.9888(2) & NPRG\\
\end{tabular}
\end{ruledtabular}
\end{table*}

For $N\ge12$ the four-loop term is the smallest of the four known terms, and the literal rule selects the three-loop sum
$S_3=1+3/(2M)$. This is less accurate than $S_2$: the deviations are $-3.4\%$, $-1.9\%$, $-1.2\%$, and $-0.4\%$ for $N=12,20,32,100$,
against $+1.0\%$, $+0.7\%$, $+0.4\%$, and $+0.1\%$ for $S_2$. The error estimate fails as well: at $N=20$ the rule gives
$\nu^{-1}=1.0833\pm0.0093$, while the NPRG value is $1.0628$. The reason is condition (ii) of Sec.~\ref{sec:ot}. $T_4$ is small for
$N\ge12$ because $c_3$ vanishes at $N=32.74$, where the three powers of $1/M$ in Eq.~(\ref{nu}) cancel. It is not the minimum of a
factorially growing sequence and does not measure the remainder. At four loops the position of the true smallest term is
not determined for $N\ge12$, and we quote both sums. The two-loop sum~(\ref{nuOT}) stays within $1\%$ of the reference for
all $N\ge6$, and the reference value of $\nu^{-1}$ lies inside $S_2\pm T_3$ for every $N\ge4$ in Table~\ref{tabnu}. We stress that for $N\ge12$ the
preference for $S_2$ rests on this comparison and on the large-$N$ argument of Sec.~\ref{sec:largeN}; it does not follow from the rule alone.

\section{The exponent \texorpdfstring{$\eta$}{eta}}\label{sec:eta}
The four-loop series reads \cite{BW,Kleinert}
\begin{align}
\eta&=\frac{\ep}{M}-\frac{(N-1)\ep^2}{M^2}+\frac{N(N-1)\ep^3}{2M^3}\nonumber\\
&\quad-\frac{(N-1)\,Q\,\ep^4}{4M^4}+O(\ep^5),\label{eta}\\
Q&=-6+2N+N^2+(-12+N+N^2)\zeta(3).\nonumber
\end{align}
It alternates in sign for all $N>2$. Its terms at $\ep=1$ are given in Table~\ref{tabterms} and shown in Fig.~\ref{figterms}(b), and its partial sums in Fig.~\ref{figpartial}(b). Unlike those of $\nu^{-1}$ in Fig.~\ref{figterms}(a), they do not decrease; the results are collected in Table~\ref{tabeta}.
The direct sum is useless: the four-loop sum is negative for every $N$ ($-0.79$, $-0.21$, $-0.055$, $-0.012$ at $N=4,6,10,20$), and
the partial sums oscillate about the true value without settling.
For $N=3,4,5$ the first term is the smallest, and the rule leaves nothing to sum; this is the situation met for the exponent
$\omega$ of the $\phi^4$ model in Ref.~\cite{S751}. For $N\ge6$ the smallest term is the third, and the truncated sum is
\begin{equation}
S_2=\frac{1}{M}-\frac{N-1}{M^2}=-\frac{1}{(N-2)^2},
\end{equation}
which is negative. Its error band $\pm T_3$ contains the reference value, e.g., $-0.0625\pm0.234$ against $0.03157(14)$ at
$N=6$, but it is many times wider than $\eta$ itself. The optimal truncation of the $2+\ep$ series therefore gives no
prediction for $\eta$ in three dimensions. Since $|T_2|<|T_1|$ requires $\ep<(N-2)/(N-1)$, condition (i) of
Sec.~\ref{sec:ot} holds only for $d<3$.

\begin{table*}[tbp]
\caption{Exponent $\eta$ in $d=3$, grouped by model as in Table~\ref{tabnu}. \emph{Nonlinear sigma model:} optimal truncation of the four-loop $2+\ep$ series with its error band (``None'' means that the first term is the smallest, so that the rule keeps no term);
bracketing midpoint, Eq.~(\ref{mid}); ${}_2F_0$ and ${}_2F_1$ resummations of Sec.~\ref{sec:hm}; and the qubit-regularized model \cite{Singh}. \emph{$\phi^4$ model} and \emph{Reference:} as in Table~\ref{tabnu}.}
\label{tabeta}
\begin{ruledtabular}
\setlength{\tabcolsep}{3.5pt}
\begin{tabular}{rccccccccc}
 & \multicolumn{5}{c}{$O(N)$ nonlinear sigma model} & \multicolumn{2}{c}{$O(N)$ $\phi^4$ model} & \multicolumn{2}{c}{Reference}\\
\cline{2-6}\cline{7-8}\cline{9-10}
 & \multicolumn{4}{c}{$2+\ep$ expansion (four loops)} & Lattice & \multicolumn{2}{c}{$4-\ve$ expansion (seven loops)} & & \\
$N$ & Opt.\ trunc. & Midpoint & ${}_2F_0$ & ${}_2F_1$ & Qubit MC & Opt.\ trunc. & Resummed & Value & Method\\
\hline
4 & None & & 0.0806 & 0.0464 & 0.0380(20) & 0.0382 & 0.0363(10) & 0.03624(8) & MC\\
5 & None & & 0.0584 & 0.0363 & & 0.0369 & 0.034591(55) & 0.03397(9) & MC\\
6 & $-0.0625\pm0.2344$ & 0.0547 & 0.0477 & 0.0324 & 0.0280(40) & 0.0353 & & 0.03157(14) & MC\\
8 & $-0.0278\pm0.1296$ & 0.0370 & 0.0354 & 0.0265 & 0.0250(20) & 0.0320 & & 0.02675(15) & MC\\
10 & $-0.0156\pm0.0879$ & 0.0283 & 0.0282 & 0.0222 & & 0.0290 & 0.0239(20) & 0.02302(12) & MC\\
12 & $-0.0100\pm0.0660$ & 0.0230 & 0.0235 & 0.0190 & & 0.0205 & & 0.0199(3) & MC\\
20 & $-0.0031\pm0.0326$ & 0.0132 & 0.0140 & 0.0119 & & 0.0137 & 0.01319(38) & 0.0129(3) & NPRG\\
32 & $-0.0011\pm0.0184$ & 0.0081 & 0.0087 & 0.0076 & & 0.0086 & & 0.00814 & SC\\
100 & $-0.0001\pm0.0053$ & 0.0025 & 0.0028 & 0.0025 & & & 0.00273(13) & 0.00268(4) & NPRG\\
\end{tabular}
\end{ruledtabular}
\end{table*}

Because the series alternates and its terms are of comparable size, a natural substitute is the mean of the two consecutive
partial sums on either side of the smallest term,
\begin{equation}
\eta_{\rm mid}=S_2+\tfrac12T_3=\frac{N^2-5N+8}{4(N-2)^3}.
\label{mid}
\end{equation}
This is not the optimal truncation rule, and we list it separately. It deviates from the reference by $+73\%$, $+38\%$, $+23\%$, and $+16\%$ at
$N=6,8,10,12$, and by $+2\%$, $-1\%$, and $-6\%$ at $N=20,32,100$. The agreement near $N=20$--$32$ is partly a crossing: for
$N\to\infty$ Eq.~(\ref{mid}) tends to $0.25/N$, while the exact result is $8/(3\pi^2N)=0.2702/N$.

\section{The large-\texorpdfstring{$N$}{N} view and the large-order behavior}\label{sec:largeN}
\begin{figure*}[tbp]\centering
\includegraphics[width=0.86\textwidth]{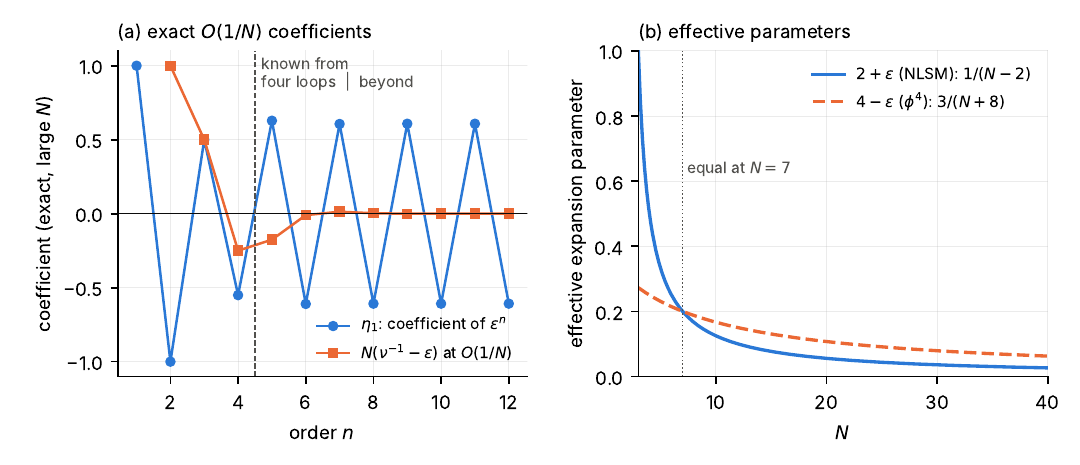}
\caption{(a) Exact $O(1/N)$ coefficients of the $2+\ep$ expansion, Eqs.~(\ref{eta1ser}) and (\ref{nu1ser}), against the order $n$; the dashed line separates the four
orders known from the four-loop series from those that follow from the $1/N$ expansion alone. The coefficients of $\eta_1$ settle at $\pm0.608$, the signature of a radius of
convergence equal to $1$, while those of $\nu^{-1}$ decay to zero, as for a radius of convergence $3$. (b) Effective expansion parameters of the two series as functions of $N$:
$\ep/(N-2)$ for the $2+\ep$ expansion and $3\ve/(N+8)$ for the $4-\ve$ expansion, at $\ep=\ve=1$. They are equal at $N=7$; for larger $N$ the $2+\ep$ parameter is the smaller one.}
\label{figlargeN}
\end{figure*}

The different behavior of the two exponents can be understood from the exact $O(1/N)$ results, valid for all $d=2\mu$ \cite{MZJ,VPK},
\begin{align}
\eta&=\frac{\eta_1}{N}+O(N^{-2}),\nonumber\\
\eta_1&=\frac{-4\,\Gamma(2\mu-2)}{\Gamma(2-\mu)\Gamma(\mu-1)\Gamma(\mu-2)\Gamma(\mu+1)},\\
\nu^{-1}&=\ep+\frac{2(\mu-1)(2\mu-1)}{2-\mu}\,\frac{\eta_1}{N}+O(N^{-2}).\label{f1}
\end{align}
Expanding numerically in $\ep=2\mu-2$ we find
\begin{align}
\eta_1&=\ep-\ep^2+\tfrac12\ep^3-0.5505\,\ep^4+0.6284\,\ep^5\nonumber\\&\quad-0.6101\,\ep^6+0.6064\,\ep^7-\dots,\label{eta1ser}\\
N(\nu^{-1}-\ep)&=\ep^2+\tfrac12\ep^3-\tfrac14\ep^4-0.1755\,\ep^5\nonumber\\&\quad-0.0098\,\ep^6+0.0134\,\ep^7+\dots\label{nu1ser}
\end{align}
The first four terms of each reproduce the $1/N$ parts of Eqs.~(\ref{eta}) and (\ref{nu}), with $-[1+\zeta(3)]/4=-0.5505$. This
consistency between the $2+\ep$ and $1/N$ expansions was used in Ref.~\cite{BW} to fix the four-loop $\beta$ function, and in Ref.~\cite{KSV} to reconstruct the
five-loop renormalization group functions. The latter are known only up to three undetermined constants, which affect the coefficients beyond the leading orders in $1/N$; we therefore restrict the analysis to four loops. The only use we make of them is as a test of the resummations in Sec.~\ref{sec:hm}, through the five-loop coefficient of $\eta$ for $N\ge8$, where the undetermined constant enters only at order $1/(N-2)^4$ and is negligible. Equations~(\ref{eta1ser}) and (\ref{nu1ser}) give the
$O(1/N)$ parts of all higher coefficients; in particular the $O(1/N)$ part of the five-loop coefficient of $\nu^{-1}$ is $-0.1755/N$. The coefficients are plotted in Fig.~\ref{figlargeN}(a).

For $\eta$ the nearest singularity of $\eta_1$ is the pole of $\Gamma(2\mu-2)$ at $\ep=-1$. The $O(1/N)$ series has unit radius of
convergence, and $\ep=1$ lies on its boundary: the terms tend to $\pm0.61/N$ and never decrease. There is no smallest term to
truncate at, and condition (i) fails for every $N$ at $d=3$. This is a property of the expansion about two
dimensions and not of the low order.

For $\nu^{-1}$ the factor $(2\mu-1)$ in Eq.~(\ref{f1}) cancels the pole at $\ep=-1$, the next pole is cancelled by zeros of the inverse
gamma functions, and the nearest singularity is at $\ep=-3$. The $O(1/N)$ series converges at $\ep=1$, to $32/(3\pi^2)=1.0808$,
with partial sums $1$, $1.5$, $1.25$, $1.0745$, $1.0647$, and $1.0781$. At this order in $1/N$ the notion of optimal truncation therefore does not
apply to $\nu^{-1}$: more terms eventually help. Within four loops the two-loop sum is the closest
($1$ against $1.0808$) and the three-loop sum the worst ($1.5$), which is the pattern seen in Table~\ref{tabnu} for $N\ge12$;
the exact five-loop $O(1/N)$ term would bring the coefficient to $1.0745$. The divergent behavior that makes the rule appropriate
at $N\le11$ sits at higher orders in $1/N$: the $\ep^4$ term alone contributes $7.91/M^2$, whereas the full $1/M^2$ coefficient of
$\nu^{-1}$ in $d=3$, from the known $O(1/N^2)$ exponent \cite{MZJ,Singh}, is $0.886$.

\subsection{The next order in \texorpdfstring{$1/N$}{1/N}}\label{sec:N2}
The $1/N$ expansion is known well beyond the leading order. In arbitrary dimension $\eta$ has been computed to $O(1/N^3)$ \cite{VPH1,VPK} and $\nu$ to $O(1/N^2)$
\cite{VPH2}; compact forms of both are given in Ref.~\cite{BGK} (see also the review \cite{GraceyRev}). These $d$-dimensional exponents describe the common large-$N$ critical point of the two models: expanded about $d=2$ they reproduce the $2+\ep$
series of the NLSM, and about $d=4$ the $4-\ve$ series of the $\phi^4$ model \cite{BGK}. They therefore presuppose universality order by order in $1/N$, and we use them only as exact
information on the $2+\ep$ coefficients, never as input at finite $N$. We write
\begin{equation}
\eta=\frac{\eta_1}{N}+\frac{\eta_2}{N^2}+\frac{\eta_3}{N^3}+\dots,\qquad
\nu^{-1}=\ep+\frac{f_1}{N}+\frac{f_2}{N^2}+\dots,
\end{equation}
with $f_1$ given by Eq.~(\ref{f1}), and use the $d$-dimensional expressions of Ref.~\cite{BGK} for $\eta_2$ and $f_2=2\lambda_2$ [Eqs.~(61)--(67) there, with $d=2-2\hat\ep$, i.e.\
$\hat\ep=-\ep/2$]. We have checked them against the known three-dimensional values, $\eta_2=-512/(27\pi^4)$ and $f_2=3.0476$, which follows from
$\nu=1-32/(3\pi^2N)+[3584/(27\pi^4)-32/\pi^2]/N^2$ \cite{MZJ}, and against $\eta_1$ and $f_1$ at several $d$. Their Taylor coefficients in $\ep$ are listed in
Table~\ref{tabN2}:
\begin{align}
\eta_2&=2\ep-3\ep^2+\tfrac52\ep^3-4.6541\,\ep^4+6.0036\,\ep^5-\dots,\label{eta2ser}\\
f_2&=2\ep^2+\ep^3+7.4093\,\ep^4-11.0193\,\ep^5+\dots\label{f2ser}
\end{align}
Three consistency checks follow. (a) The first four coefficients of each reproduce the $1/N^2$ parts of the four-loop series~(\ref{nu}) and (\ref{eta}) once these are
re-expanded in $1/N$; for instance $7.4093/N^2$ is the $1/N^2$ part of $c_3$, which in powers of $1/M$ becomes the $7.9093/M^2$ quoted above. (b) The $\ep^5$
coefficient $-11.0193$ coincides with the $1/N^2$ part of the five-loop coefficient of $\nu^{-1}$ of Ref.~\cite{KSV}, whose undetermined constants enter only
from $O(1/N^3)$ on; the same holds for the $\ep^5$ coefficient of $\eta_2$. (c) The $O(1/N^3)$ result $\eta_3=4\ep-8\ep^2+9\ep^3-\tfrac14[40+27\zeta(3)]\ep^4+17.184\,\ep^5+\dots$ [Eq.~(70) of Ref.~\cite{BGK}] reproduces the
$1/N^3$ parts of all four coefficients of $\eta$, with $-[40+27\zeta(3)]/4=-18.114$, and its $\ep^5$ term reproduces the $1/N^3$ part of the five-loop coefficient of Ref.~\cite{KSV}.

\begin{table}[tbp]
\caption{Taylor coefficients of the exact $1/N$ expansion in $\ep=d-2$: $\eta=\sum_k\eta_k/N^k$ and $\nu^{-1}=\ep+\sum_kf_k/N^k$. Orders $n\le4$ are contained in the
four-loop series; the last row gives the functions at $\ep=1$.}
\label{tabN2}
\begin{ruledtabular}
\begin{tabular}{rrrrr}
$n$ & $\eta_1$ & $\eta_2$ & $f_1$ & $f_2$ \\ \hline
1 & $1$ & $2$ & $0$ & $0$\\
2 & $-1$ & $-3$ & $1$ & $2$\\
3 & $0.5$ & $2.5$ & $0.5$ & $1$\\
4 & $-0.5505$ & $-4.6541$ & $-0.25$ & $7.4093$\\
5 & $0.6285$ & $6.0036$ & $-0.1755$ & $-11.0193$\\
6 & $-0.6101$ & $-7.2361$ & $-0.0098$ & $13.6367$\\
7 & $0.6064$ & $9.5388$ & $0.0134$ & $-26.1900$\\
8 & $-0.6080$ & $-11.7511$ & $0.0030$ & $36.7993$\\ \hline
$\ep=1$ & $0.2702$ & $-0.1947$ & $1.0808$ & $3.0476$
\end{tabular}
\end{ruledtabular}
\end{table}

The new information concerns the analytic structure. At $O(1/N^2)$ the cancellation that removes the pole of $f_1$ at $\ep=-1$ does not occur: both $\eta_2$ and $f_2$ are singular there, through the double pole of the squared gamma-function prefactor and the poles of the polygamma functions $\psi(1+\ep)$ and $\psi'(1+\ep)$, with
poles of order three ($\eta_2$) and four ($f_2$). The radius of convergence is $1$ for both, and the point $d=3$ again lies on the circle of convergence.
The coefficients alternate in sign (from $n=4$ on for $f_2$) and grow like $n^2$ and $n^3$ respectively, i.e.\ geometrically times a power, not factorially. At $\ep=1$ the partial sums
of $f_2$ are $2$, $3$, $10.41$, $-0.61$, $13.03$, $-13.16$ through $\ep^7$, and those of $\eta_2$ are $2$, $-1$, $1.5$, $-3.15$, $2.85$, $-4.39$, while the exact values
are $3.0476$ and $-0.1947$. The convergence of $\nu^{-1}$ found at $O(1/N)$ is therefore a property of the leading order only. The growing terms that make optimal
truncation the appropriate rule for $N\le11$ begin at $O(1/N^2)$, and the first of them is precisely the four-loop coefficient $7.41\,\ep^4/N^2$.
The power of $n$ also increases from one order in $1/N$ to the next ($n^0$, $n^2$ for $\eta$), so the large-order behavior at fixed $N$ cannot be read off term by term.

The dependence of the accuracy on $N$ has a simple origin. Every coefficient beyond the leading one in Eqs.~(\ref{nu}) and (\ref{eta}) carries
at least one power of $1/(N-2)$, so that the series is effectively an expansion in $\ep/(N-2)$. This parameter decreases as $N$ grows,
just as $\sigma\ve=3\ve/(N+8)$ does for the $\phi^4$ model \cite{S751,S1049} [Fig.~\ref{figlargeN}(b)]; the two parameters are equal at $N=7$, and for larger $N$ that of the $2+\ep$ series is the smaller one, and the error estimate of the two-loop truncation, $1/[2(N-2)]$, decreases with it.
In the limit $N\to\infty$ all corrections vanish order by order and the perturbation series by itself reproduces the exact large-$N$ exponents,
$\nu=1/(d-2)$ and $\eta=0$, for every truncation. For $\nu^{-1}$ the statement extends to the next order: since the $O(1/N)$ series~(\ref{nu1ser}) converges at $\ep=1$,
the partial sums approach the exact coefficient $32/(3\pi^2)$ as more loops are included. For $\eta$ it does not, because the $O(1/N)$ series~(\ref{eta1ser}) is at the
boundary of its circle of convergence; the exact coefficient $8/(3\pi^2)$ is then recovered only after resummation (Sec.~\ref{sec:hm}).
At small $N$ the situation is the opposite. The parameter $\ep/(N-2)$ is of order one ($1$ at $N=3$, $1/2$ at $N=4$), the smallest term is large, and only four
terms are available. The truncation error is then intrinsically large, and four coefficients are too few for a resummation to compensate:
the seven-loop $4-\ve$ series, with a smaller parameter and almost twice as many terms, is accurate already at $N=4$. The limited accuracy at small $N$ thus reflects the short series and the
large effective parameter, and need not signal a failure of the expansion itself.

Little is known about the large-order behavior of the $2+\ep$ series itself. The absence of sign alternation in $\nu^{-1}$ led to the
suggestion that terms of the form $\exp(-C/\ep)$ are needed \cite{HB}. For the two-dimensional model, the weak-coupling series of
the ground-state energy has positive coefficients for $N=3$ and $4$ and is not Borel summable, which is attributed to infrared renormalons and is expected at any finite $N$;
in the large-$N$ limit the infrared renormalon poles are absent, and the leading ultraviolet renormalon gives a Borel-summable, sign-oscillating series. An additional singularity in the Borel plane,
at a position proportional to $N-2$, corresponds to the instanton \cite{Volin}. Later work has shown that the renormalon structure is more subtle, and that large-$N$ estimates of the renormalon positions can be misleading \cite{MMR}.
Our observations show a similar pattern: positive, growing coefficients of $\nu^{-1}$
at small $N$, a natural parameter $\ep/(N-2)$, and a convergent $O(1/N)$ part.

These results do not carry over directly to the exponents, because they concern an observable of the two-dimensional model, expanded in the coupling and obtained from its exact
Bethe-ansatz solution, whereas the exponents are series in $\ep$ determined by the renormalization group functions at the fixed point, for which no exact solution is available in $d=2+\ep$.
Moreover, in minimal subtraction the renormalization group functions are generally expected to be free of renormalons, their large orders being governed by instantons, as in the $\phi^4$ model.
The instanton singularity found in Ref.~\cite{Volin} nevertheless suggests a simple picture, which we state as a conjecture. If the renormalization group functions inherit a Borel singularity at a position
proportional to $N-2$, the coefficients of the $\ep$ series grow as $n!\,[c\,(N-2)]^{-n}$, with $c$ a constant. Such a term is of order $N^{-n}$, so that at any fixed order in $1/N$ it contributes to a finite number of
coefficients only: the factorial growth is invisible in the $1/N$ expansion, in accord with the purely geometric growth found at $O(1/N)$ and $O(1/N^2)$ in Sec.~\ref{sec:N2}. At finite $N$ it would overtake the
geometric growth at orders $n$ of the order of $N-2$. With four known terms the series would then be in the factorial regime for $N=3$--$5$, where the terms of $\nu^{-1}$ already grow and optimal truncation is the appropriate rule, and in the
geometric regime at large $N$. This would explain why both divergent-type and convergent-type approximants are of use in Sec.~\ref{sec:hm}, the latter being the better choice at large $N$.
A derivation of the large-order behavior of the
critical exponents in $2+\ep$ dimensions, e.g.\ by a Lipatov-type instanton calculation, is, to our knowledge, not available.

\section{Hypergeometric-Meijer resummation}\label{sec:hm}
To see what a resummation of the same four coefficients adds, we apply the hypergeometric-Meijer (HM) algorithm \cite{HM1,HM2} at its lowest
orders. The idea is to replace the truncated series by a hypergeometric function whose Taylor coefficients reproduce the known ones and whose
analytic structure, a power-law singularity or a factorial growth controlled by a single parameter $\sigma$, mimics that of the unknown function.
Writing either series as $c_0\,\ep^{p}\sum_n r_n\ep^n$ with $r_0=1$, we use
\begin{equation}
{}_pF_q(\{a_i\};\{b_j\};\sigma\ep)=\sum_{n=0}^\infty\frac{(a_1)_n\cdots(a_p)_n}{(b_1)_n\cdots(b_q)_n}\,\frac{(\sigma\ep)^n}{n!},
\label{pFq}
\end{equation}
whose coefficient ratios are rational in $n$,
\begin{equation}
n\,\frac{r_n}{r_{n-1}}\equiv nR_n=\sigma\,\frac{\prod_{i=1}^{p}(n-1+a_i)}{\prod_{j=1}^{q}(n-1+b_j)}.
\label{ratioeq}
\end{equation}
Equating the right-hand side to the known ratios gives equations that are linear in the elementary symmetric functions of the $a_i$ once the denominator is
multiplied out, so that the fit reduces to a small linear system and the $a_i$ follow as roots of a polynomial. For ${}_2F_0$ ($p=2$, $q=0$) the three ratios
$R_1$, $R_2$, $R_3$ fix $\sigma$, $a_1+a_2$, and $a_1a_2$; one finds $\sigma=(3R_3-4R_2+R_1)/2$. The function is evaluated at $\ep=1$
through its Mellin-Barnes (Meijer-$G$) representation, which is valid also when the series diverges (Appendix~\ref{app:mb}). For $\nu^{-1}$ the fit gives
\begin{equation}
\sigma=\frac{-7N^2+60N-110+54(N-3)\zeta(3)}{4(N-2)^2},
\end{equation}
which is positive for $N<14.93$. The singularity then lies on the positive axis, in line with the non-Borel character of the series, and we take the
median of the two lateral sums; their imaginary part gives an ambiguity of about $\pm0.07$ in $\nu$ for $N=4$--$12$. For $N>14.93$,
$\sigma<0$ and the sum is unique. For $\eta$ the fit gives $\sigma<0$ for all $N$ (from $-1.34$ at $N=4$ to $-1.19$ at $N=100$, tending to $-[1+3\zeta(3)]/4$), and the sum is unique.

The results are listed in Tables~\ref{tabnu} and \ref{tabeta}. For $\nu$ the resummation deviates from the reference by $-9.4\%$, $-3.1\%$, $-0.9\%$,
$-0.3\%$, and $-1.1\%$ for $N=4,5,6,8,10$, the same quality as the two-loop truncation~(\ref{nuOT}); it does not repair $N=4,5$. For
$N=12$ and $20$ it is $2.6$--$2.7\%$ low, because $\sigma$ passes through zero at $N=14.93$ and the approximant degenerates there; it recovers
to $-1.2\%$ and $-0.3\%$ at $N=32$ and $100$. Its large-$N$ limit gives $N(\nu^{-1}-1)\to1.387$, against the exact $1.0808$ and against $1$ for
Eq.~(\ref{nuOT}). This is as expected from Sec.~\ref{sec:largeN}: a factorially divergent ansatz is imposed on a series whose leading large-$N$
part converges.

For $\eta$ the resummation is a clear improvement over the direct sum. The resummed values
are positive and decrease smoothly with $N$. They exceed the reference by $122\%$, $72\%$, $51\%$, $32\%$, $23\%$, and $18\%$ for $N=4,5,6,8,10,12$, and by
$9\%$, $7\%$, and $4\%$ for $N=20,32,100$; the large-$N$ limit is $N\eta\to0.278$ against the exact $0.2702$. Going one order higher in the
approximant does not change this: a ${}_3F_1$ fitted to $1+\eta$ with $\sigma$ fixed at its large-$N$ value lowers $\eta$ by at most $5\%$
(0.0767 at $N=4$) and by less than $2\%$ for $N\ge20$. The resummed $2+\ep$ series thus gives
the correct $\eta$ at large $N$ and the correct order of magnitude at $N\approx10$, but overestimates it by a factor of two at $N=4$.

\begin{table*}[tbp]
\caption{Parameters of the hypergeometric approximants. Columns 2--4: ${}_2F_0(a_1,a_2;;\sigma\ep)$ fitted to $\nu^{-1}/\ep$; columns 5--7: ${}_2F_0$ fitted to $\eta/\ep$;
columns 8--11: ${}_2F_1(a_1,a_2;b;\sigma\ep)$ fitted to $1+\eta$ (Sec.~\ref{sec:hm}). A pair of complex conjugate parameters is shown in one cell.
For $\nu^{-1}$, $\sigma>0$ for $N<14.93$; for $\eta$, $\sigma<0$ for every $N$ in both approximants.}
\label{tabpar}
\begin{ruledtabular}
\begin{tabular}{rcccccccccc}
 & \multicolumn{3}{c}{$\nu^{-1}$: ${}_2F_0$} & \multicolumn{3}{c}{$\eta$: ${}_2F_0$} & \multicolumn{4}{c}{$\eta$: ${}_2F_1$}\\
\cline{2-4}\cline{5-7}\cline{8-11}
$N$ & $\sigma$ & $a_1$ & $a_2$ & $\sigma$ & $a_1$ & $a_2$ & $\sigma$ & $a_1$ & $a_2$ & $b$\\
\hline
4 & $+5.182$ & $-0.780$ & $-0.124$ & $-1.339$ & \multicolumn{2}{c}{$-0.313\pm1.011i$} & $-2.768$ & $-1.753$ & $-0.163$ & $-1.582$\\
6 & $+3.011$ & $-0.616$ & $-0.135$ & $-1.564$ & \multicolumn{2}{c}{$-0.420\pm0.789i$} & $-2.451$ & $-1.904$ & $-0.096$ & $-1.801$\\
8 & $+1.712$ & \multicolumn{2}{c}{$-0.257\pm0.177i$} & $-1.533$ & \multicolumn{2}{c}{$-0.446\pm0.750i$} & $-2.257$ & $-1.921$ & $-0.070$ & $-1.828$\\
10 & $+0.955$ & \multicolumn{2}{c}{$-0.042\pm0.359i$} & $-1.486$ & \multicolumn{2}{c}{$-0.458\pm0.740i$} & $-2.140$ & $-1.923$ & $-0.056$ & $-1.829$\\
12 & $+0.465$ & $0.412$ & $0.521$ & $-1.444$ & \multicolumn{2}{c}{$-0.465\pm0.738i$} & $-2.062$ & $-1.921$ & $-0.046$ & $-1.824$\\
20 & $-0.468$ & $-3.057$ & $0.039$ & $-1.343$ & \multicolumn{2}{c}{$-0.479\pm0.746i$} & $-1.905$ & $-1.911$ & $-0.027$ & $-1.799$\\
100 & $-1.505$ & $-1.662$ & $0.004$ & $-1.193$ & \multicolumn{2}{c}{$-0.496\pm0.775i$} & $-1.712$ & $-1.887$ & $-0.006$ & $-1.747$\\
\end{tabular}
\end{ruledtabular}
\end{table*}

The fitted parameters are collected in Table~\ref{tabpar}. They vary smoothly with $N$ except for the ${}_2F_0$ of $\nu^{-1}$ near $N=14.93$, where $\sigma$
changes sign and the approximant degenerates; we return to this below.

\subsection{Convergent-type approximants}
The ${}_pF_{p-2}$ family used above presupposes Gevrey-1 behavior, i.e.\ coefficients growing factorially, $c_n\sim n!\,\sigma^n$. This is the
established large-order behavior of the $4-\ve$ series, but for the $2+\ep$ series the large-order behavior is not known \cite{HB}, and
factorial growth is not guaranteed. To our knowledge the radius of convergence of the finite-$N$ $2+\ep$ series has not been determined; it is usually taken to be zero \cite{HB,Kleinert}, but this has not been shown; the poor agreement of the resummed four-loop series at $N=3$ noted in Ref.~\cite{DCR} is an empirical observation, not a statement about the large-order behavior. The only higher-order information available points the other way. At $O(1/N)$ the coefficients of
$\eta$ are those of Eq.~(\ref{eta1ser}); their ratios approach $-1$ ($-0.971$, $-0.994$, $-1.003$, $-1.000$ for $n=5$ to $8$), which is geometric growth with
radius of convergence $1$, while Gevrey-1 growth would make the ratios grow linearly with $n$. The same holds for $\nu^{-1}$, with radius $3$. At finite $N$
the three known ratios of the $\eta$ series show no growth with $n$ either (for example $-1.10$, $-0.60$, $-1.40$ at $N=12$).
This motivates the convergent-type ${}_pF_{p-1}$ approximants, whose coefficients grow geometrically, $c_n\sim n^{\gamma}\sigma^n$. The choice rests on this analytic information alone, which is available before any comparison with three-dimensional data, and not on the agreement with the reference values.

The lowest approximant of this type with a free parameter $\sigma$ is the four-parameter ${}_2F_1(a_1,a_2;b;\sigma\ep)$. Equation~(\ref{ratioeq}) with $q=1$, multiplied by $(n-1+b)$, is linear in $b$, $\sigma$, $\sigma(a_1+a_2)$, and $\sigma a_1a_2$, and four ratios fix all four. We fit it to the four ratios of $1+\eta$, the same device as for
the ${}_3F_1$ above: adding a constant is the simplest way to obtain a fourth ratio from four coefficients, and the constant was fixed at $1$ in advance (its influence is examined below). We then evaluate the approximant at $\ep=1$; $\sigma$ comes out negative for all $N$ ($-2.77$ at $N=4$, tending to $-1.66$ for $N\to\infty$), so the sum is
unique, and the fitted parameters vary smoothly with $N$ without approaching poles ($-1.83\le b\le-1.58$). The results are given in Table~\ref{tabeta}. The improvement over ${}_2F_0$ at
intermediate $N$ is large: the deviations from the reference are $+28\%$, $+6.8\%$, $+2.5\%$, $-0.9\%$, $-3.6\%$, and $-4.7\%$ for $N=4,5,6,8,10,12$, against
$+122\%$, $+72\%$, $+51\%$, $+32\%$, $+23\%$, and $+18\%$ for ${}_2F_0$. At large $N$ the ${}_2F_1$ settles about $8\%$ below the reference ($-7.9\%$, $-7.0\%$, $-7.9\%$ at $N=20,32,100$), with
$N\eta\to0.249$ against the exact $0.2702$. The constant added to $\eta$ is arbitrary; replacing $1+\eta$ by $\lambda+\eta$ with $1/4\le\lambda\le4$ changes the result by less
than $4\%$ for $N\ge8$ and by less than $1\%$ for $N\ge20$.

The reason for the improvement can be seen in the coefficients each approximant predicts beyond the known ones. At $O(1/N)$ the ${}_2F_1$ predicts $0.684$ and
$-0.905$ for the coefficients of $\ep^5$ and $\ep^6$, against the exact $0.628$ and $-0.610$, whereas the ${}_2F_0$ predicts $1.089$ and $-3.23$, the factorial growth it
imposes. At finite $N$, the five-loop coefficient of $\eta$ is known from Ref.~\cite{KSV} up to a constant that enters only at order $1/(N-2)^4$ and is negligible for $N\ge8$;
the ${}_2F_1$ predicts it within $41\%$, $24\%$, and $11\%$ at $N=12,20,100$, the ${}_2F_0$ within $128\%$, $100\%$, and $78\%$. By building in factorial growth, the ${}_2F_0$
overestimates the higher-order terms and hence $\eta$; the ${}_2F_1$, with geometric growth, follows the actual coefficients more closely. At large $N$ neither
approximant reproduces the exact $O(1/N)$ coefficient, since $\ep=1$ lies on the boundary of the circle of convergence of $\eta_1$: the ${}_2F_0$ gives $N\eta\to0.278$, above the
exact $0.2702$, and the ${}_2F_1$ gives $0.249$, below it. The ${}_2F_0$ overestimates $\eta$ because it builds in factorial growth of the coefficients; the ${}_2F_1$ underestimates it slightly.
As a consequence, for every $N\ge8$ the ${}_2F_0$ and ${}_2F_1$ values bracket the reference value of $\eta$ (Table~\ref{tabeta}).

For $\nu^{-1}$ the convergent-type approximants do not help. A ${}_2F_1$ with $\sigma=-1/3$, the position of the exact $O(1/N)$ singularity, or with $\sigma$ free fitted to $1+\nu^{-1}$,
varies erratically with $N$: $\sigma$ changes sign at $N=12$ and $20$, and the result at $N=10$ is $9.5\%$ low. The series of $\nu^{-1}$ is positive and dominated by its first two terms,
and four coefficients leave little for such an approximant to add; for $\nu$ the two-loop truncation, Eq.~(\ref{nuOT}), remains the best estimate from the $2+\ep$ side.

\emph{Test at higher orders in $1/N$.} The exact results of Sec.~\ref{sec:N2} allow a test of the approximants that goes beyond the leading large-$N$ limit.
Each approximant is a function of $N$ at $d=3$; expanding it in $1/N$ (numerically, from its values for $200\le N\le16000$) and comparing with the exact coefficients
tests how well it extrapolates in $\ep$ at each order in $1/N$ separately. The result is given in Table~\ref{tabNexp}. For $\eta$ the ${}_2F_1$ reproduces the
$1/N^2$ coefficient within $6\%$ ($-0.184$ against $-0.195$), although the four coefficients it is fitted to contain only the first four terms of
Eq.~(\ref{eta2ser}), whose partial sum at $\ep=1$ is $-3.15$; it also has the right sign at $O(1/N^3)$. The ${}_2F_0$ is closer at $O(1/N)$ but has the wrong sign
at $O(1/N^2)$, which is why it overestimates $\eta$ at intermediate $N$. For $\nu$ the two-loop truncation, $\nu=1-1/N-1/N^2-\dots$, is the closest of the three at both
orders; its $1/N^2$ coefficient is off by about $50\%$, but at $N=20$ this amounts to $0.2\%$ of $\nu$. The four-loop sum and the ${}_2F_0$ overshoot at both orders.

\begin{table}[tbp]
\caption{Coefficients of the $1/N$ expansion in $d=3$ of the approximants, $\nu=1+\nu_1/N+\nu_2/N^2+\dots$ and $\eta=\eta_1/N+\eta_2/N^2+\eta_3/N^3+\dots$,
against the exact values \cite{MZJ,BGK}.}
\label{tabNexp}
\begin{ruledtabular}
\setlength{\tabcolsep}{3pt}
\begin{tabular}{lrrrrr}
 & $\nu_1$ & $\nu_2$ & $\eta_1$ & $\eta_2$ & $\eta_3$\\ \hline
Exact & $-1.0808$ & $-1.8796$ & $0.2702$ & $-0.1947$ & $-1.8812$\\
4-loop sum & $-1.25$ & $-8.847$ & $-0.0505$ & $-3.154$ & $-13.11$\\
2-loop OT & $-1$ & $-1$ & -- & -- & --\\
${}_2F_0$ & $-1.387$ & $-3.346$ & $0.2780$ & $0.0502$ & $-0.039$\\
${}_2F_1$ & -- & -- & $0.2487$ & $-0.1836$ & $-0.687$
\end{tabular}
\end{ruledtabular}
\end{table}

\subsection{Choice of the expanded quantity: \texorpdfstring{$1/\eta$ and $\nu$}{1/eta and nu}}\label{sec:var}
The analysis so far expands $\nu^{-1}$ and $\eta$, the quantities the renormalization group functions give directly. With four terms the choice of variable
matters, and we have examined the two alternatives, $1/\eta$ and $\nu$. The results are summarized in Table~\ref{tabvar}.

\emph{The series of $1/\eta$.} Inverting Eq.~(\ref{eta}) gives
\begin{equation}
\frac{\ep}{\eta}=(N-2)+(N-1)\,\ep+\tfrac12(N-1)\,\ep^2+g_3\,\ep^3+O(\ep^4),
\label{inveta}
\end{equation}
with $g_3=2.928$, $4.536$, $6.884$, $12.44$, $56.50$ at $N=4,6,10,20,100$. All coefficients are positive, so the alternation of the $\eta$ series disappears and the partial
sums of $\eta$ decrease monotonically toward the reference; but the terms still do not decrease at $\ep=1$ ($g_1>g_0$), and the four-loop sum overestimates $\eta$ by
$193\%$, $53\%$, and $23\%$ at $N=4,10,100$. The exact large-$N$ results explain both features. From $1/\eta=N/\eta_1-\eta_2/\eta_1^2+O(1/N)$, the leading part
$\ep/\eta_1=1+\ep+\tfrac12\ep^2+0.5505\,\ep^3+0.2226\,\ep^4+0.2042\,\ep^5+\dots$ is singular only at the zeros of $\eta_1$, a double zero at $\ep=2$ ($d=4$) and a simple zero at
$\ep=-2$. Its radius of convergence is therefore $2$, and at $\ep=1$ it converges to $3\pi^2/8=3.701$, but slowly: the four-loop sum is $3.05$. The next part,
$-\ep\,\eta_2/\eta_1^2=-2-\ep-\tfrac12\ep^2+1.452\,\ep^3+\dots$, has a simple pole at $\ep=-1$ and radius $1$; its four-loop sum is $-2.05$, against the exact $8/3$.
Inversion thus moves the difficulty from the leading to the subleading order in $1/N$ without removing it.

The radius $2$ of the leading part suggests a convergent-type ${}_2F_1$ with $\sigma=+1/2$ fixed, fitted to the three ratios of Eq.~(\ref{inveta}). Applied to the
four exact coefficients of $\ep/\eta_1$ it gives $4.26$ (exact $3.70$), i.e.\ $N\eta\to0.235$. At finite $N$ it gives the best small-$N$ values of the paper:
$+16\%$, $+3.0\%$, and $-2.4\%$ at $N=4,5,6$, against $+28\%$, $+6.8\%$, and $+2.5\%$ for the ${}_2F_1$ on $1+\eta$. For $N\ge8$ it settles $7$--$14\%$ below the
reference, and the ${}_2F_1$ on $1+\eta$ is better. A ${}_2F_0$ with free $\sigma$ fails for $1/\eta$ ($\sigma>0$, errors of $64$--$423\%$), as expected for a
non-alternating series. The two ${}_2F_1$ estimates, from $1+\eta$ and from $1/\eta$, differ by at most $7.6\%$ for $6\le N\le100$; we regard this spread as a
measure of the uncertainty of the four-loop resummation of $\eta$.

\emph{The series of $\nu$.} Inverting Eq.~(\ref{nu}) gives
\begin{equation}
\ep\,\nu=1-\frac{\ep}{M}+\Big(\frac{1}{M^2}-\frac{1}{2M}\Big)\ep^2+h_3\,\ep^3+O(\ep^4).
\label{nuser}
\end{equation}
Here the inversion does not help. The two-loop sum is $\nu=(N-3)/(N-2)$, and optimal truncation gives $-14.6\%$, $-7.2\%$, $-2.1\%$, and $-0.5\%$ at $N=5,6,8,10$,
against $-3.9\%$, $-1.0\%$, $+0.7\%$, and $+1.0\%$ for Eq.~(\ref{nuOT}); at $N=4$ the $\ep^2$ coefficient vanishes and condition (ii) of Sec.~\ref{sec:ot} fails.
For $N\ge12$ the rule selects the three-loop sum, with errors of $-4.5\%$ to $-0.4\%$. A ${}_2F_0$ fit is no better ($-13.6\%$ at $N=5$, $-2.4\%$ at $N=10$), and a
${}_2F_1$ with $\sigma=-1/3$, the position of the $O(1/N)$ singularity, is erratic at intermediate $N$. The large-$N$ structure is the same as for $\nu^{-1}$: the $O(1/N)$
part of Eq.~(\ref{nuser}), $-f_1(\ep)/\ep$, converges with radius $3$, while the $O(1/N^2)$ part, $(f_1^2/\ep-f_2)/\ep$, has radius $1$ and coefficients $-2$, $0$,
$-6.41$, $10.77$, $-14.24$, \dots, against the exact sum $-1.880$. The difference lies in how the two variables reorganize the series: Eq.~(\ref{nuOT}),
$\nu=1-1/N-1/N^2-1/N^3-\dots$, has coefficients of order one at every order in $1/N$, while the two-loop sum of $\nu$, $1-1/(N-2)=1-1/N-2/N^2-4/N^3-\dots$,
has coefficients that double from one order to the next. For $\nu$ the natural variable is therefore $\nu^{-1}$, the quantity given directly by the
derivative of the $\beta$ function.

\begin{table}[tbp]
\caption{Alternative expanded quantities in $d=3$. Columns 2--3: $\eta$ from the four-loop sum of $1/\eta$, Eq.~(\ref{inveta}), and from its ${}_2F_1$ resummation with
$\sigma=1/2$. Columns 4--5: $\nu$ from the series of $\nu$, Eq.~(\ref{nuser}): the optimally truncated sum (OT; two-loop sum for $N\le10$, three-loop sum for $N\ge12$) and the ${}_2F_0$ resummation.
Deviations from the reference values of Tables~\ref{tabnu} and \ref{tabeta} in percent in parentheses.}
\label{tabvar}
\begin{ruledtabular}
\setlength{\tabcolsep}{1pt}\footnotesize
\begin{tabular}{rcccc}
 & \multicolumn{2}{c}{$\eta$ from $1/\eta$} & \multicolumn{2}{c}{$\nu$ from $\nu$}\\
\cline{2-3}\cline{4-5}
$N$ & Sum & ${}_2F_1$ & OT & ${}_2F_0$\\ \hline
4 & 0.1061 (+193) & 0.0421 (+16) & 0.5000 ($-33$) & --\\
5 & 0.0778 (+129) & 0.0350 (+3.0) & 0.6667 ($-15$) & 0.6738 ($-14$)\\
6 & 0.0624 (+98) & 0.0308 ($-2.4$) & 0.7500 ($-7.2$) & 0.7556 ($-6.5$)\\
8 & 0.0450 (+68) & 0.0248 ($-7.3$) & 0.8333 ($-2.1$) & 0.8289 ($-2.6$)\\
10 & 0.0352 (+53) & 0.0206 ($-10$) & 0.8750 ($-0.5$) & 0.8588 ($-2.4$)\\
12 & 0.0290 (+46) & 0.0176 ($-11$) & 0.8600 ($-4.5$) & 0.8691 ($-3.5$)\\
20 & 0.0170 (+32) & 0.0111 ($-14$) & 0.9198 ($-2.2$) & 0.9138 ($-2.9$)\\
32 & 0.0105 (+29) & 0.0071 ($-13$) & 0.9511 ($-1.3$) & 0.9517 ($-1.3$)\\
100 & 0.0033 (+23) & 0.0023 ($-13$) & 0.9848 ($-0.4$) & 0.9857 ($-0.3$)
\end{tabular}
\end{ruledtabular}
\end{table}

\section{Comparison of the two models}\label{sec:comp}
\begin{figure*}[tbp]\centering
\includegraphics[width=0.86\textwidth]{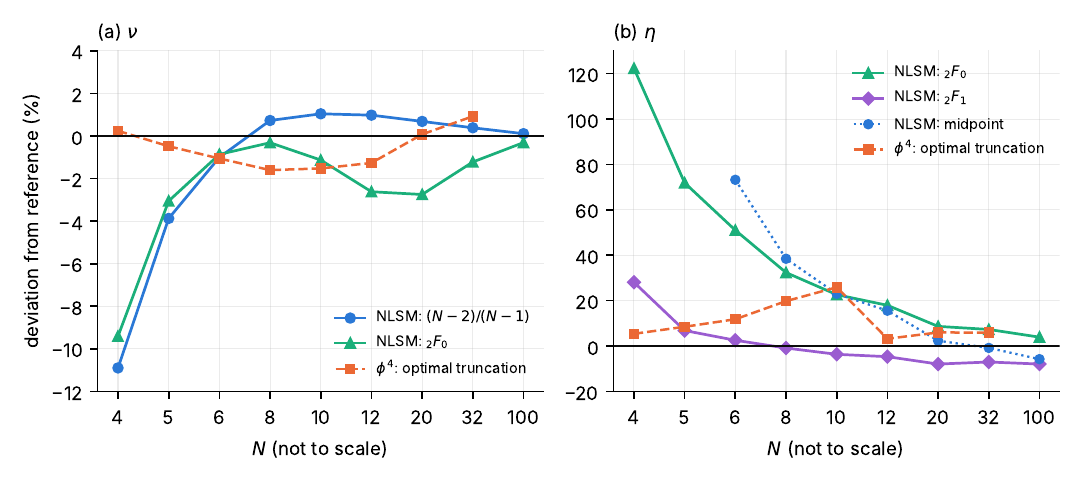}
\caption{Deviation from the reference values of Tables~\ref{tabnu} and \ref{tabeta}, in percent, against $N$ (the horizontal axis lists the values of $N$ studied and is not to scale).
(a) $\nu$: the two-loop optimal truncation of the $2+\ep$ series, its ${}_2F_0$ resummation, and the two-loop optimal truncation of the seven-loop $4-\ve$ series.
(b) $\eta$: the ${}_2F_0$ and ${}_2F_1$ resummations and the bracketing midpoint of the $2+\ep$ series, and the optimal truncation of the $4-\ve$ series. In both panels the
deviations of the $2+\ep$ results shrink as $N$ grows; for $\eta$ the ${}_2F_0$ and ${}_2F_1$ values lie on opposite sides of the reference for $N\ge8$.}
\label{figdev}
\end{figure*}
Figure~\ref{figdev} summarizes the accuracy of all the estimates as a function of $N$. For $\nu$ [Fig.~\ref{figdev}(a)], the two-loop truncations of the two expansions
lie within $\pm1.6\%$ of the reference for every $N\ge6$, and the $2+\ep$ error falls to $0.1\%$ at $N=100$. For $\eta$ [Fig.~\ref{figdev}(b)], the ${}_2F_0$ error decreases
monotonically, from $122\%$ at $N=4$ to $4\%$ at $N=100$, while the ${}_2F_1$ stays within $\pm8\%$ for $N\ge5$. For $5\le N\le10$ the ${}_2F_1$ estimate of $\eta$ from the four-loop
$2+\ep$ series is closer to the reference than the optimal truncation of the seven-loop $4-\ve$ series ($6.8\%$, $2.5\%$, $0.9\%$, and $3.6\%$ against $8.5\%$, $11.8\%$, $19.8\%$, and $25.9\%$ in magnitude);
the seven-loop resummation \cite{S1049} is more accurate at $N=4$ and $5$ ($0.2\%$ and $1.8\%$) and comparable at $N=10$ ($3.8\%$).

\begin{figure*}[tbp]\centering
\includegraphics[width=0.86\textwidth]{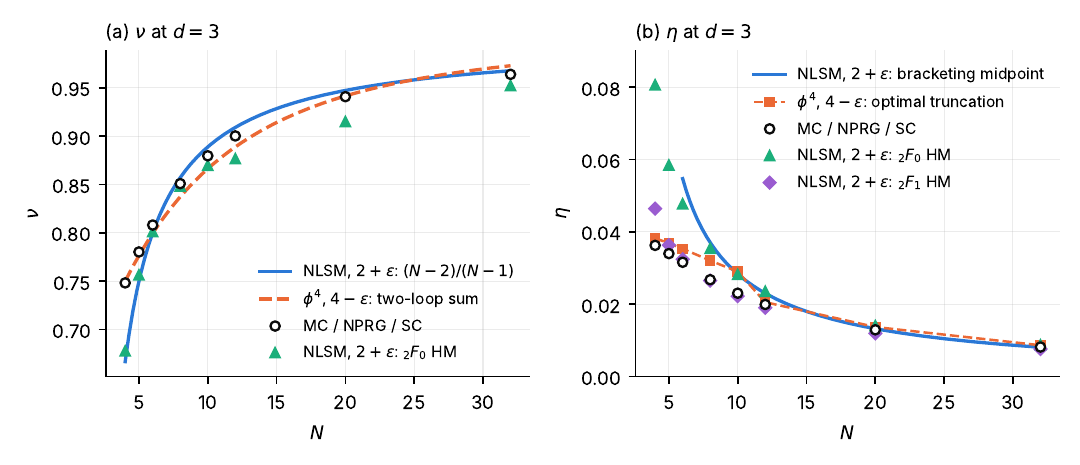}
\caption{(a) $\nu$ in $d=3$ from the two-loop truncation of the $2+\ep$ series, Eq.~(\ref{nuOT}), and from the two-loop optimal truncation of the $4-\ve$ series \cite{S751},
together with the ${}_2F_0$ resummation of Sec.~\ref{sec:hm} and the reference values of Table~\ref{tabnu}. (b) $\eta$ in $d=3$ from the bracketing midpoint, Eq.~(\ref{mid}),
from the ${}_2F_0$ and ${}_2F_1$ resummations, and from the optimal truncation of the $4-\ve$ series \cite{S751}.}
\label{fig}
\end{figure*}
For the $\phi^4$ model the optimal truncation of $\nu^{-1}$ is the two-loop sum up to $N=65$ \cite{S751},
\begin{equation}
\nu^{-1}\simeq 2-\frac{N+2}{N+8}-\frac{(N+2)(13N+44)}{2(N+8)^3}.
\label{phi4OT}
\end{equation}
It deviates from the reference by $+0.2\%$, $-0.5\%$, $-1.1\%$, $-1.6\%$,
$-1.5\%$, $-1.3\%$, $+0.1\%$, and $+0.9\%$ for $N=4,5,6,8,10,12,20,32$.

Both expansions thus reduce, at their optimal order, to a two-loop rational function of $N$, one obtained from $d=4$ and
one from $d=2$. Their relative difference is $-11.1\%$ at $N=4$, $-3.4\%$ at $N=5$, $+0.06\%$ at $N=6$, $+2.4\%$, $+2.6\%$, and $+2.3\%$
at $N=8,10,12$, and $+0.6\%$ and $-0.5\%$ at $N=20,32$ [Fig.~\ref{fig}(a)]. For $N=8,10,12$ the two truncations lie on opposite sides of
the MC value; this is not the case for $N=5,6,20,32$, so that no bracketing property can be claimed. For $N\ge6$ each truncation
agrees with the nonperturbative results at the $1$--$2\%$ level. At $N=4$ and $5$ the $\phi^4$ truncation remains accurate while the NLSM one does not, in line with the size of the
respective parameters, $3/(N+8)$ against $1/(N-2)$.

It is instructive to compare with the nonperturbative NLSM results of Ref.~\cite{Singh}, $\nu=0.751(5)$, $0.804(4)$, and $0.835(6)$ for $N=4,6,8$.
Equation~(\ref{nuOT}) gives $0.667$, $0.800$, and $0.857$: it agrees with the qubit model at $N=6$, is $2.6\%$ above it at $N=8$, and fails at $N=4$.
For $N=6$ the older MC value $0.8180(50)$ \cite{HS} is higher than both the qubit result and the recent MC value $0.8081(6)$ \cite{Has};
the two truncations, $0.800$ and $0.7996$, favor the lower values.

Resummation does not change this picture. The seven-loop resummation of the $4-\ve$ series \cite{S1049}
agrees with MC within its quoted uncertainty for $N=4,5,10$, whereas the ${}_2F_0$ resummation of the $2+\ep$ series is no closer to the
reference than Eq.~(\ref{nuOT}). The variational interpolation of Ref.~\cite{Kleinert} reaches the correct values at small $N$ only by using the $4-\ve$
coefficients as input.

For $\eta$ the comparison is one sided. The optimal truncation of the $4-\ve$ series gives $\eta$ within $3$--$12\%$ of the reference for $N=4,5,6,12,20,32$ and within
$20$--$26\%$ for $N=8,10$ \cite{S751}, and the resummed series within about $4\%$ for $N=4,5,10$ \cite{S1049}. The $2+\ep$ series offers no optimal truncation at
$d=3$, the midpoint estimate~(\ref{mid}) and the ${}_2F_0$ resummation are quantitative only for $N\ge20$, while the ${}_2F_1$ resummation is within $5\%$ for $6\le N\le12$ and within $8\%$ beyond [Fig.~\ref{fig}(b)].

These findings bear on the question raised in Refs.~\cite{FWZ,DM,DCR}. Agreement of a single exponent at the percent level cannot establish that
the two fixed points are the same conformal field theory; it shows only that the $2+\ep$ expansion, taken by itself, is numerically
compatible with the WF value of $\nu$ for $N\ge6$. Conversely, the failure at $N=3$--$5$ and for $\eta$ cannot be read as evidence against
universality: at $O(1/N)$, where the two expansions are known to describe the same function of $d$, the $2+\ep$ series of $\eta$ already has no decreasing terms at $\ep=1$.
The region $N=3,4$, where a distinct fixed point has been suggested \cite{DCR}, is precisely where four-loop $2+\ep$ data are uninformative. For these two values a recent study of the operator spectrum, based on the one-loop dimensions of the operators that would have to recombine with the protected one, concludes that the NLSM and WF fixed points are most likely distinct theories, while for $N>4$ a coincidence of the two in $d=3$ alone is not excluded \cite{DCR2}; this analysis concerns the operator content and not the exponents studied here.

\section{Discussion}\label{sec:disc}
\emph{What the comparison shows.} The universality hypothesis predicts that the $2+\ep$ expansion of the NLSM and the $4-\ve$ expansion of the $\phi^4$ model,
continued to $d=3$, describe the same exponents. The test performed here is internal to perturbative renormalization group theory: each series is evaluated
separately, with the same rule, and the results are compared with each other and with nonperturbative data. For $\nu$ the two expansions reduce, at their optimal
order, to two different rational functions of $N$, obtained from opposite ends of the interval $2<d<4$, which agree within $2.6\%$ for every $N\ge6$ and within $0.6\%$ at $N=20$ and $32$.
For $\eta$ the agreement requires resummation of the $2+\ep$ series but is then of comparable quality at large $N$. This is the content of universality that four loops
can test, and the test is passed in the large-$N$ regime.

\emph{Why large $N$.} Two independent facts single out large $N$. First, the effective parameter $\ep/(N-2)$ decreases with $N$ and, for $N>7$, is smaller than that of
the $4-\ve$ series. Second, at $O(1/N)$ the $2+\ep$ series is convergent, with radius $3$ for $\nu^{-1}$ and $1$ for $\eta$, so that its leading large-$N$ part is fully
controlled. The poor behavior at small $N$ comes from the higher orders in $1/N$, which are large ($7.91/M^2$ from the four-loop term alone, against $0.886/M^2$ for the exact
sum). The exact $O(1/N^2)$ exponents show why: their $\ep$ expansions have unit radius of convergence, with poles of order three and four at $d=1$, and their
coefficients grow like $n^2$ and $n^3$, so that already at this order the terms at $\ep=1$ increase. They also show that the ${}_2F_1$ extrapolates correctly at
$O(1/N^2)$ (Table~\ref{tabNexp}).

\emph{What is free of large-$N$ input.} Since the large-$N$ information enters only through the four-loop $\beta$ function, the two exponents differ in this respect. For $\nu$ nothing beyond three loops is needed (Sec.~\ref{sec:nu}). For $\eta$ we have checked whether the three directly computed coefficients suffice, by resumming them alone (Table~\ref{tab3loop}). A ${}_1F_0$ fitted to $\eta/\ep$ gives the closed form $\eta=(N-2)^{N-2}/(N-1)^{N-1}$, and a $[1/1]$ Pad\'e approximant gives $\eta=1/(3N-4)$; a ${}_2F_1$ fitted to $1+\eta$ with $\sigma=-1$, the position of the exact $O(1/N)$ singularity, and a ${}_2F_0$ fitted to $1+\eta$ complete the set. All overestimate $\eta$, by $16$--$310\%$ for the convergent-type forms and by more for the ${}_2F_0$, whose $\sigma$ comes out positive for every $N$; their large-$N$ limits, $N\eta\to1/e$, $1/3$, $0.306$, and $0.558$, all exceed the exact $0.2702$. The four-loop coefficient is thus indispensable for $\eta$, and the accuracy of the four-loop ${}_2F_1$ cannot be separated from the large-$N$ input contained in it.

\begin{table}[tbp]
\caption{$\eta$ in $d=3$ from the three directly computed coefficients of the $2+\ep$ series alone, compared with the four-loop ${}_2F_1$ of Table~\ref{tabeta}. Deviations from the reference values in percent; last row: the large-$N$ limit of $N\eta$ (exact $0.2702$). For the ${}_2F_0$ the median of the lateral sums is used ($\sigma>0$).}
\label{tab3loop}
\begin{ruledtabular}
\setlength{\tabcolsep}{2pt}\footnotesize
\begin{tabular}{rccccc}
 & \multicolumn{4}{c}{three loops} & four loops\\
\cline{2-5}\cline{6-6}
$N$ & ${}_1F_0$ & Pad\'e $[1/1]$ & ${}_2F_1$, $\sigma=-1$ & ${}_2F_0$ & ${}_2F_1$\\ \hline
4 & $+309$ & $+245$ & $+265$ & $+1368$ & $+28$\\
6 & $+159$ & $+126$ & $+121$ & $+505$ & $+2.5$\\
8 & $+112$ & $+87$ & $+78$ & $+326$ & $-0.9$\\
12 & $+76$ & $+57$ & $+47$ & $+216$ & $-4.7$\\
20 & $+54$ & $+38$ & $+29$ & $+156$ & $-7.9$\\
100 & $+39$ & $+26$ & $+16$ & $+115$ & $-7.9$\\
$N\eta$, $N\to\infty$ & 0.368 & 0.333 & 0.306 & 0.558 & 0.249\\
\end{tabular}
\end{ruledtabular}
\end{table}

\emph{Limitations.} The four-loop coefficients of the $2+\ep$ series were not obtained from diagrams alone. The terms through three loops follow from direct calculations \cite{BZJ,HB}, and so does the four-loop field anomalous dimension, but the four-loop $\beta$ function was determined using, in addition, the exponent $\eta$ to $O(1/N^3)$ \cite{BW}; a later direct four-loop calculation for the Grassmannian models gives the $O(N)$ result only through isomorphisms and $1/N$ information \cite{Wegner89}. That the higher orders are not fully independent of large-$N$ input has also been stressed in Ref.~\cite{DCR}. This concerns the $\ep^4$ coefficients only: it does not affect $\nu$ (Sec.~\ref{sec:nu}), but it does affect $\eta$, for which the agreement we find at large $N$ is therefore partly built into the coefficients. What the present analysis tests is how far the truncated series, taken at fixed finite $N$, reproduces the three-dimensional values. Moreover, agreement of two exponents at the percent level does not prove that the two fixed points are the same conformal field theory \cite{DCR}; it shows
that the perturbative data from both sides are consistent with it. Our error estimates are those of the optimal truncation rule or the spread between approximants; they are
not rigorous bounds. The choice between the two- and three-loop sums for $12\le N\le32$ is not fixed by the rule at four loops. The ${}_2F_1$ resummation of $\eta$ relies on
the device of fitting $1+\eta$ to obtain a fourth ratio, whose effect we have checked to be small for $N\ge8$. Finally, the exponents themselves are scheme
independent, but the large-order behavior of the $2+\ep$ series, which would decide between divergent- and convergent-type approximants, has not been determined.

\emph{Outlook.} The most useful new input would be the complete five-loop renormalization group functions of the NLSM, which would fix the truncation order for
$N\ge12$ and add a ratio to every resummation. On the $1/N$ side, $\eta_3$ is known in closed form in arbitrary dimension \cite{VPK}; here we have used only its
Taylor coefficients through $\ep^5$ and its value in $d=3$, and its full $\ep$ dependence would test the approximants at third order in $1/N$, as would $\nu$ at $O(1/N^3)$, which to our knowledge has not been computed in arbitrary dimension.

\section{Conclusions}\label{sec:concl}
The few orders known for the renormalization group functions of the $O(N)$ nonlinear sigma model are usually considered insufficient to
study its universality with the $O(N)$ vector $\phi^4$ model. Our main result is that at large $N$ they are sufficient. Treated
independently, with the optimal truncation rule that works for the $4-\ve$ expansion \cite{S751} and with hypergeometric-Meijer resummation,
the four-loop $2+\ep$ series gives accurate exponents in three dimensions, in agreement with the $\phi^4$ model and with nonperturbative results.
To our knowledge this is the first comparison of the two models carried out entirely within perturbative renormalization group theory,
with the two expansions never combined; as noted in Sec.~\ref{sec:disc}, the four-loop $2+\ep$ coefficients themselves contain large-$N$ input. In detail:

(i) The $2+\ep$ series is effectively an expansion in $\ep/(N-2)$. As for the $4-\ve$ series, whose parameter $3/(N+8)$ decreases with $N$ \cite{S1049}, the accuracy is controlled by $N$; for $N>7$ the $2+\ep$ parameter is the smaller of the two. It decreases with $N$, the error estimate of the truncation
decreases with it, and for $N\to\infty$ the series reproduces the exact large-$N$ exponents order by order.

(ii) For $\nu$ the series by itself is accurate. Its optimal truncation for $3\le N\le11$ is the two-loop sum, $\nu=(N-2)/(N-1)$, which agrees
with MC and NPRG values within about $1\%$ for all $N\ge6$ and within $0.1\%$ at $N=100$, and with the optimally truncated $4-\ve$ series within $2.6\%$.
For $N\ge12$ the literal rule selects the three-loop sum, because the four-loop coefficient passes through zero near $N\simeq33$; the two-loop
sum remains the more accurate one, but the rule alone does not fix the truncation order there at four loops. The two-loop result and its three-loop error estimate use only directly computed coefficients; the agreement for $\nu$ is thus independent of the large-$N$ information that enters the $2+\ep$ series at four loops.

(iii) For $\eta$ the raw series fails at every $N$: its terms do not decrease at $\ep=1$, because the exact $O(1/N)$ series has unit radius of
convergence in $\ep$. Hypergeometric-Meijer resummation turns the negative direct sum into a positive, smooth function of $N$. Because the large-order behavior of the $2+\ep$ series is unknown, Gevrey-1 growth cannot be assumed, and the convergent-type ${}_2F_1$, motivated by the geometric growth of the exact $O(1/N)$ coefficients, is within $5\%$ of the reference for $6\le N\le12$; it also predicts the next coefficients better than the divergent-type ${}_2F_0$, and its $1/N$ expansion in $d=3$ reproduces the exact $O(1/N^2)$ coefficient of $\eta$ within $6\%$, where the ${}_2F_0$ has the wrong sign. The ${}_2F_0$ error falls steadily with $N$,
from $51\%$ at $N=6$ to $9\%$ at $N=20$ and $4\%$ at $N=100$, and for every $N\ge8$ the two approximants bracket the reference value; their large-$N$ limits, $N\eta\to0.278$ and $0.249$, bracket the exact $0.2702$. Expanding $1/\eta$ instead removes the sign alternation, and its ${}_2F_1$ resummation gives the best values for $4\le N\le6$ ($3\%$ at $N=5$); the spread between the two ${}_2F_1$ estimates, at most $7.6\%$ for $N\ge6$, measures the uncertainty. For $\nu$ resummation,
of either type, confirms the truncated result without improving on it, and expanding $\nu$ instead of $\nu^{-1}$ makes the results worse.

(iv) For small $N$ the effective parameter $\ep/(N-2)$ is of order one and only four terms are known. The NLSM results are then poor
($-10.9\%$ for $\nu$ at $N=4$; $\eta$ too large by $28\%$ with ${}_2F_1$ and by a factor of two with ${}_2F_0$), while the seven-loop $\phi^4$ series is accurate from $N=4$.

The expansions about two and four dimensions therefore agree in the large-$N$ regime, which is consistent, within perturbative renormalization
group theory, with the universality of the two models there. For $\nu$ the test is independent of the $1/N$ expansion, since only the directly computed low-order coefficients of each series enter; for $\eta$, which needs the four-loop coefficient and with it large-$N$ input, the agreement is a quantitative test of the truncated series at finite $N$ rather than independent evidence for universality. For small $N$, where the analytic connection of the two fixed points has been
questioned \cite{DCR}, four loops do not allow a conclusion; higher orders, or resummations that use more of the known analytic structure,
would be needed.

\section*{Declaration of generative AI and AI-assisted technologies in the writing process}

The author acknowledges the use of the AI tool Claude (Anthropic; models Claude Opus 5.5 and Claude Fable 5.2) in the preparation of this work. Under the author's direction, the tool was used to search and summarize the literature, to adapt and extend the author's Mathematica code for the hypergeometric-Meijer resummation and to translate it into Python, to generate the figures and tables, and to assist in drafting and revising the text. The author reviewed and edited all AI-assisted content and takes full responsibility for the content of this publication.

\appendix
\section{Mellin-Barnes evaluation of the approximants}\label{app:mb}
\begin{figure}[!htbp]\centering
\includegraphics[width=\columnwidth]{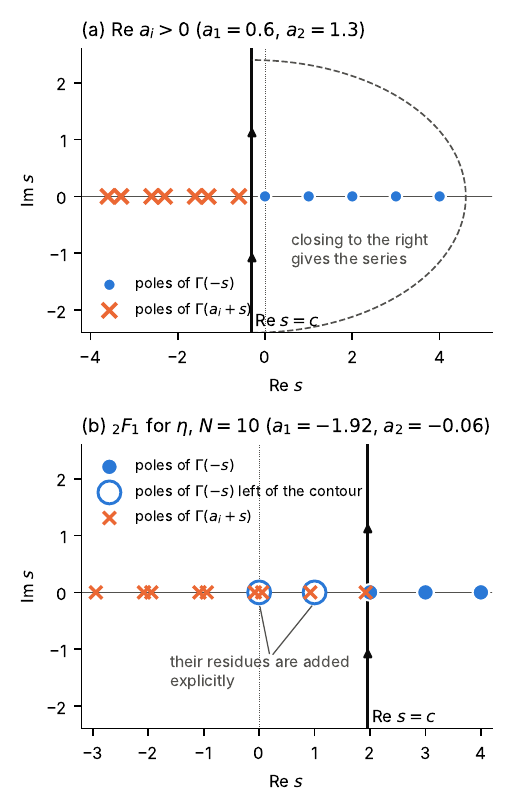}
\caption{Contour of the Mellin-Barnes integral~(\ref{mb}) in the complex $s$ plane. Dots: poles of $\Gamma(-s)$ at $s=0,1,2,\dots$; crosses: poles of
$\Gamma(a_i+s)$ at $s=-a_i-n$. Top: for $\mathrm{Re}\,a_i>0$ a vertical line $\mathrm{Re}\,s=c$ with $-\min\mathrm{Re}\,a_i<c<0$ separates the two sets; closing the
contour to the right (dashed) picks up the residues of $\Gamma(-s)$ and reproduces the series~(\ref{pFq}). Bottom: the ${}_2F_1$ fitted to $1+\eta$ at $N=10$
(Sec.~\ref{sec:hm}), with $a_1=-1.92$ and $a_2=-0.06$: the poles of $\Gamma(a_i+s)$ extend to $s=1.92$, so the line is placed at $c=1.96$, and the poles of
$\Gamma(-s)$ at $s=0$ and $1$, now to the left of it (open circles), are added explicitly as the first two terms of the series.}
\label{figmb}
\end{figure}

For $x>0$ the hypergeometric function (\ref{pFq}) with $p\le q+2$ has the Mellin-Barnes representation
\begin{align}
{}_pF_q(a;b;-x)&=\frac{\prod_j\Gamma(b_j)}{\prod_i\Gamma(a_i)}\,\frac{1}{2\pi i}\int_{c-i\infty}^{c+i\infty}\!ds\nonumber\\
&\quad\times\Gamma(-s)\,\frac{\prod_i\Gamma(a_i+s)}{\prod_j\Gamma(b_j+s)}\,x^{s},
\label{mb}
\end{align}
where the contour separates the poles of $\Gamma(-s)$ at $s=0,1,2,\dots$ from those of $\Gamma(a_i+s)$ at $s=-a_i-n$. Closing the contour to the right reproduces the series~(\ref{pFq}),
term by term, whenever it converges. For $p=q+2$, as for ${}_2F_0$, the series diverges, but the integral is well defined and equals its Borel sum. For $\sigma<0$ the argument
$\sigma\ep$ is negative, $x=|\sigma|\ep$, and Eq.~(\ref{mb}) gives a unique real value. For $\sigma>0$ the argument is positive; we continue $x\to x\,e^{\pm i\pi}$, which
gives the two lateral sums. Since the fitted parameters are real or form complex-conjugate pairs, the two lateral sums are complex conjugates of each other: their mean is real and is the median value quoted in the text, and half their difference, which is imaginary, measures the ambiguity.
In practice we take $c$ between the right-most pole of $\Gamma(a_i+s)$ and the next integer (Fig.~\ref{figmb}). When some $a_i$ have negative real part, so that $c>0$, the residues of
$\Gamma(-s)$ at the integers below $c$ are added explicitly as the first terms of the series. On the line $s=c+it$ the integrand decreases, up to a power of $|t|$, as
$\exp[-(1+p-q)\pi|t|/2+|\arg x|\,|t|]$: as $e^{-\pi|t|/2}$ for the lateral sums of ${}_2F_0$ and as $e^{-\pi|t|}$ for ${}_2F_1$ at negative argument, so that the
integral is evaluated to high accuracy by standard quadrature. We checked the procedure against closed forms: for ${}_2F_0$ at negative argument against
$x^{-a_1}U(a_1,1+a_1-a_2,1/x)$, with $U$ Tricomi's function, for the lateral sums of ${}_2F_0$ against the same expression continued to $x\,e^{\pm i\pi}$, with $U$ computed from its connection formula in terms of the two convergent Kummer series,
and for ${}_2F_1$ against the Gauss function; for all the approximants used in this paper the agreement is better than $10^{-12}$.

\FloatBarrier

\end{document}